\documentclass{aa}
\usepackage{times}
\usepackage{graphics}
\usepackage{xspace}
\usepackage{epsfig}
\usepackage{jwaabib}
\usepackage{rotating}
\usepackage{dcolumn}

\newcommand{\myrule}{\rule[-0.2cm]{0.cm}{0.4cm}}

\begin{document}

\title{{\em XMM-Newton} probes the stellar population in Cha\,I South}

\author{B. Stelzer\inst {1} \and G. Micela\inst {1} \and R. Neuh\"auser\inst {2} }

\institute{INAF - Osservatorio Astronomico di Palermo,
  Piazza del Parlamento 1,
  I-90134 Palermo, 
  Italy \and 
  Astrophysikalisches Institut und Universit\"ats-Sternwarte,
  Schillerg\"asschen 2-3, 
  D-07745 Jena, 
  Germany} 

\offprints{B. Stelzer}
\mail{B. Stelzer, stelzer@astropa.unipa.it}
\titlerunning{{\em XMM-Newton} observation of Cha\,I}

\date{Received $<$05-02-2004$>$ / Accepted $<$31-03-2004$>$}

\abstract{We report on a 30\,ksec {\em XMM-Newton} observation of the central region of 
the Cha\,I star forming cloud. The field includes a substantial fraction of the
known pre-main sequence population of Cha\,I South, including all thirteen known very-low
mass H$\alpha$ emitters. We detect two bona-fide brown dwarfs (spectral types M7.5 and M8)
and seven H$\alpha$ emitting objects near the hydrogen burning mass limit, including 
six of seven earlier detections by {\em ROSAT}. 
Three objects classified as Cha\,I candidate members according to their NIR photometry
are revealed by {\em XMM-Newton}, providing further evidence for them being truely young stars. 
A total of $11$ new X-ray sources without known optical/IR counterpart may comprise further
as yet unrecognized faint cloud members. 
Spectral analysis of the X-ray bright stars 
shows that previous X-ray studies in Cha\,I 
have underestimated the X-ray luminosities, as a result of simplified assumptions
on the spectral shape. In particular, the extinction is variable over the field, 
such that the choice of a uniform value for the column density is inappropriate. 
We establish that the X-ray saturation level for the late-type stars in Cha\,I is located 
near $L_{\rm x}/L_{\rm bol} \sim 10^{-2.5}$, 
with a possible decline to $L_{\rm x}/L_{\rm bol} \sim 10^{-3}$ for the lowest mass stars. 
A group of strongly absorbed stars with unusual hard X-ray emission 
is clustered around HD\,97048, a HAeBe star and the only confirmed intermediate-mass star in 
the field. While the X-ray properties of HD\,97048 are indistinguishable from its lower-mass 
neighbors, another presumably A-type star (identified as such based on NIR photometry) stands 
out as the softest X-ray emitter in the whole sample. This suggests that various X-ray emission
mechanisms may be at work in intermediate-mass pre-main sequence stars. 
We find that X-ray luminosity follows a tight correlation with age, 
effective temperature, and mass. No dramatic changes in these correlations are seen
at the substellar boundary, suggesting that the same dynamo mechanism operates in
both low-mass stars and brown dwarfs, at least at young ages. 
The variability of the lowest-mass objects is also 
similar to that of higher-mass T Tauri stars. X-ray flares are seen on about $1/10$th 
of the Cha\,I members in the field. 
}

\maketitle

\section{Introduction}\label{sect:intro}

The Chamaeleon cloud complex is one of the most nearby star forming
complexes, composed of three major clouds. 
Their isolated position at high galactic latitude   
($b \approx -15^\circ$), resulting in both low foreground extinction and
low contamination with background objects,   
makes them attractive targets for the study of the formation of low-mass stars. 

Cha\,I hosts the largest number of known low-mass pre-main sequence (PMS) stars 
and is one of the best-studied of these regions (see e.g. \cite{Schwartz77.1}, 
\cite{Gauvin92.1}, \cite{Prusti92.1}, \cite{Hartigan93.1}).  
A substantial number of the Cha\,I cloud members have first been identified
as X-ray sources in pointed {\em ROSAT} observations (\cite{Feigelson93.1};
henceforth F93), and later -- 
with help of optical observations -- been confirmed as PMS stars 
(\cite{Lawson96.1}; henceforth LFH96). 
In contrast to most of the previously known Cha\,I members, the so-called 
classical T Tauri stars (cTTS), these latter ones 
belong to the class of weak-line T Tauri stars (wTTS), i.e. they show only weak
H$\alpha$ emission, presumably because accretion has ceased after the dispersal
of the circumstellar disk. LFH96 estimate that the number ratio of wTTS to cTTS
in Cha\,I is $\geq 2$. Both populations of TTS stars are mixed in the 
Hertzsprung-Russell diagram (HRD), indicating a wide range of disk lifetimes. 

In the last few years a large number of faint new candidate members of Cha\,I have 
been proposed based on their near-infrared (NIR) colors
(\cite{Cambresy98.1}, \cite{Oasa99.1}, \cite{Persi00.1}, 
\cite{Kenyon01.1}, \cite{Gomez01.1}, \cite{Carpenter02.1}). 
Several of these candidates have been confirmed to be low-mass stars on basis
of their NIR spectra (\cite{Gomez03.1}). 

The masses of the presently known
Cha\,I members reach down into the substellar regime, including 13 very-low
mass (VLM) objects at or below the hydrogen burning mass limit 
discovered in an H$\alpha$ survey by \citey{Comeron99.1}. 
ChaH$\alpha$\,1 (spectral type M7.5; \cite{Comeron00.1}; henceforth CNK00) 
was the first brown dwarf to be detected in X-rays in a {\em ROSAT}
pointed observation (\cite{Neuhaeuser98.1}). 
However, {\em ROSAT} observations were hampered by low sensitivity and low
spatial resolution, and thus unable to constrain the X-ray properties of the
VLM H$\alpha$ sources. 

The X-ray emission of VLM stars and brown dwarfs is poorly constrained. 
In lack of sensitive observations it is unclear how far the solar-stellar
connection reaches into the VLM regime. In particular the influence of 
mass, temperature, and age on the activity of the coolest stars and the substellar
objects remains obscure. 
The only bona-fide field brown dwarf detected in X-rays
so far, LP\,944-20, is of intermediate age ($500$\,Myrs), and was revealed  
with {\em Chandra} only during a flare (\cite{Rutledge00.1}). 
A subsequent deep {\em XMM-Newton} observation
was not able to recover the source (\cite{Martin02.1}) suggesting that this object
exhibits only episodic outbursts of activity. 
The X-ray properties of VLM stars and brown dwarfs on the PMS 
may be different, because at young ages the atmospheres are hotter, 
such that more ions are present
enhancing the coupling between matter and magnetic field (\cite{Mohanty02.1}).

Recent deep {\em XMM-Newton} and {\em Chandra} observations centered on nearby 
regions of star formation such as $\rho$\,Oph (\cite{Imanishi01.1}), 
IC\,348 (\cite{Preibisch02.1}) and the Orion Nebular Cluster (\cite{Feigelson02.1}) 
started to open up the X-ray window to the brown dwarf regime. 
However, these observations provided at most a few dozen counts per source 
and/or concern a poorly characterized population of VLM objects. 
Cha\,I is one of the nearest star forming regions with the best-studied group
of young brown dwarfs known to date, 
providing the highest sensitivity yet for the detection of the lowest
mass stars and brown dwarfs at young ages. 

In this paper we present the {\em XMM-Newton} observation 
of the central region of the Cha\,I South cloud.
The observation is described in Sect.~\ref{sect:obs_and_data}, where we also
outline the steps of the data reduction. The nature of the X-ray sources is
discussed in Sect.~\ref{sect:nature}. A detailed spectral analysis is performed
for the brighter X-ray sources, and an analysis based on hardness ratios for the
fainter X-ray sources (Sect.~\ref{sect:spec}). In Sect.~\ref{sect:lcs} we present 
our variability study. The results on individual stars and groups of stars are   
summarized in Sect.~\ref{sect:indiv}. Finally, we provide a comparison with earlier
{\em ROSAT} observations of the same field (Sect.~\ref{sect:rosat}), and we examine
correlations between X-ray emission and stellar parameters (Sect.~\ref{sect:corr}).
A summary of our findings is given in Sect.~\ref{sect:summary}.

\section{Observations and Data Analysis}\label{sect:obs_and_data}

A 30\,ksec {\em XMM-Newton} observation centered on the candidate brown dwarf
ChaH$\alpha$\,3 was carried out on April 9, 2002 with the European
Photon Imaging Camera (EPIC) as prime instrument. EPIC consists of
three cameras, two EPIC-MOS detectors and one EPIC-pn detector
(see \cite{Jansen01.1}, \cite{Turner01.1}, and \cite{Strueder01.1}
for details on the instruments). 
The observations were performed in full-frame mode employing the thin filter
for all components of the EPIC. The Optical Monitor was in blocked position. 
Data from the Reflection Grating Spectrograph will not be used because 
we were interested in imaging of the star forming region.  
The field-of-view (FOV) of EPIC comprises all 13 VLM objects in the Cha\,I 
region, and also includes a number of higher-mass T Tauri members of the 
star forming complex. The observing log for all CCD instruments onboard 
{\em XMM-Newton} is given in Table~\ref{tab:obslog}.
%
%
\begin{table}
\begin{center}
\caption{Observing log for the {\em XMM-Newton} observation of Cha\,I on April 9, 2002 (Obs-ID 0002740501; Revolution number 427). "Start" and "Stop" refer to start and end of exposures.}
\label{tab:obslog}
\begin{tabular}{lrrrrr}\hline
Instr. & \multicolumn{2}{c}{UT [hh:mm:ss]} & \multicolumn{2}{c}{JD - 2452373} & Expo \\ 
      & \multicolumn{1}{c}{Start} & \multicolumn{1}{c}{Stop} & \multicolumn{1}{c}{Start} & \multicolumn{1}{c}{Stop} & \multicolumn{1}{c}{[ksec]} \\ \hline
pn     & 10:20:23 & 18:59:59 & 0.930556 & 1.290972 & 31.14 \\ 
MOS\,1 & 09:47:05 & 19:03:49 & 0.907639 & 1.293750 & 33.36 \\
MOS\,2 & 09:47:06 & 19:03:51 & 0.907639 & 1.293750 & 33.36 \\ \hline
\end{tabular}
\end{center}
\end{table}

\subsection{Data Reduction}\label{subsect:data_red}

As a first step in the data reduction we checked for times of high background.
To do so we generated a lightcurve for the whole CCD array, 
and searched for the limiting acceptable count rate that provides the highest
signal-to-noise. We screened data from the three instruments separately, and
subsequently applied the resulting good time intervals (GTIs) to the events tables. 
It turned out that about half of the observing time is contaminated by 
protons from solar flares. The useful exposure time retained for further analysis
is 15.9\,ksec for pn, 
and 20.3/20.7\,ksec for MOS\,1/MOS\,2 which are less sensitive to 
high background. The data was filtered for pixel patterns, 
edge effects at the boundary of individual CCD chips, and a few
excessively luminous columns were removed. By eliminating the lowest
pulse height channels 
we further reduced the noise. 

After the filtering process we binned images in various energy bands
and for all EPIC instruments from the cleaned event files. We used a bin size
of $5^{\prime\prime}$ making full use of the spatial resolution of EPIC. 
We limit our analysis to energies between $0.3 - 7.8$\,keV.
Below $\sim 0.3$\,keV the instrument calibration is not well understood yet. 
The upper boundary of $7.8$\,keV avoids the inclusion of additional noise,
because for high energies the image is background dominated. 
We split the broad band into three energy bands: 

\begin{tabular}{lcl}
soft    & = & 0.3\,keV $-$ 1.0\,keV \\
hard\,1 & = & 1.0\,keV $-$ 2.4\,keV \\
hard\,2 & = & 2.4\,keV $-$ 7.8\,keV \\
\end{tabular}

\subsection{Source Detection}\label{subsect:detection}

Source detection was performed in all energy bands (soft, hard\,1, hard\,2, 
and broad) with the {\em XMM-Newton} Science Analysis System (SAS) 
pipeline, version 5.4.1,  
which employs the box and the maximum likelihood detection algorithm
({\em eboxdetect} and {\em emldetect}). First the box
algorithm is run in local mode with a low detection threshold. The resulting
source list is only used to generate background maps. In the next step all
sources above a certain threshold are cut out, the resulting source-free image 
is rebinned, corrected for spatial exposure variations and smoothed 
with a two-dimensional spline fit.
For this process we eliminated all sources with likelihood $> 15$. Lower
threshold values would eliminate a major part of the image such that 
spline fitting becomes difficult. For the subsequent source detection with
{\em eboxdetect} in map mode we used a $ML$ threshold of $8$. 
Then we run {\em emldetect} on the source list produced by {\em eboxdetect} in map mode. 
For {\em emldetect} we used a $ML$ threshold of $10$. 
However, a known problem in the current version of the {\it emldetect} code leads 
to an overestimation of the $ML$. 
Following the advise in {\em XMM-Newton} News No.\,29 in the final source list we retain
only sources with $ML > 20$.  

We proceeded in two steps. 
First all EPIC instruments were analysed separately as described above, 
then the source detection procedure was applied to the combined files of
pn, MOS\,1, and MOS\,2. Before merging the data of the three detectors 
we checked the relative alignment of the images by comparing the position
of sources common to all three images. We found no systematic displacement. 
To take account of the different sensitivity of pn and MOS we computed 
the ratio of count rates measured by pn, MOS\,1, and MOS\,2
for the brighter sources detected with all three instruments during 
the first source detection step. In the broad band we found a median ratio 
pn/MOS $\approx 3.2$. The exposure map of pn was then scaled with this value 
before adding up the files.  
Differences in the spectral response of MOS and pn were taken into account 
by determining and applying a separate scaling factor for each energy band:
pn/MOS $\approx 4.0$ (soft), 2.6 (hard\,1), and 2.8 (hard\,2).
The merged observation corresponds to an equivalent MOS exposure time
of $\sim 90$\,ksec in the center of the image. 

\begin{figure*}
\begin{center}
\resizebox{16cm}{!}{\includegraphics{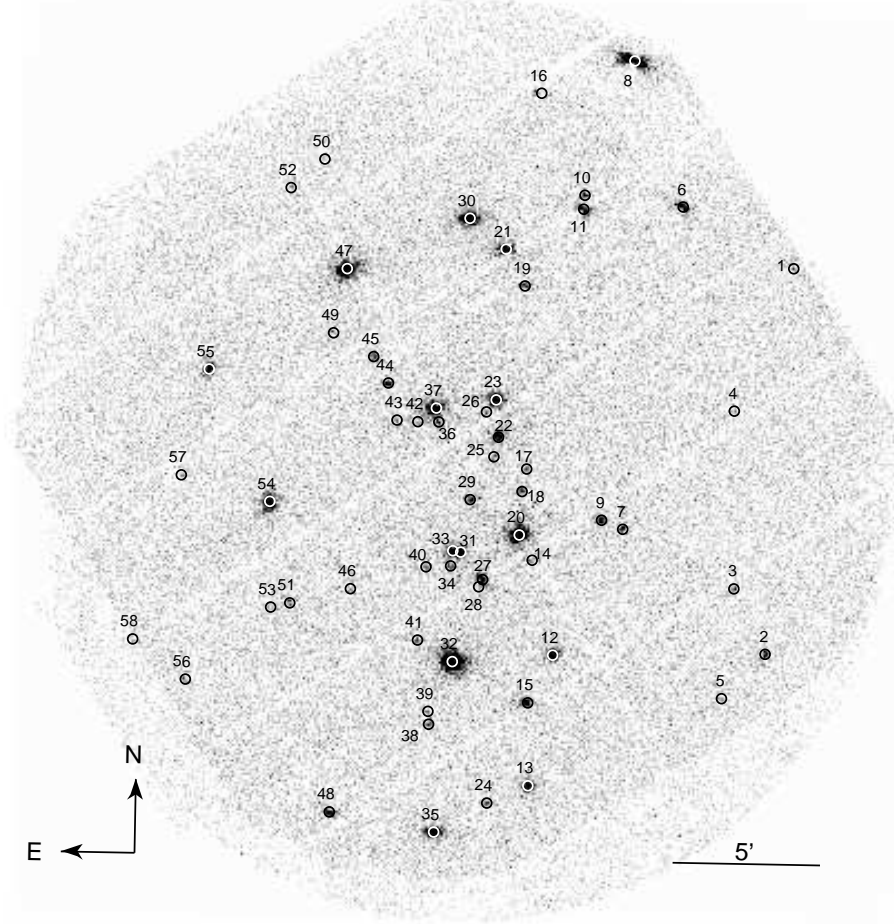}}
\caption{Merged image of EPIC pn, MOS\,1 and MOS\,2 for the center of the Cha\,I South cloud. The positions of all $58$ detected sources are marked by circles (radius=$10^{\prime\prime}$). Numbers refer to the designations in column~1 of Table~\ref{tab:x-sources}.}
\label{fig:mergedimage}
\end{center}
\end{figure*}

The merged MOS/pn data contains 58 X-ray sources with $ML > 20$. 
For further analysis we use the result from the merged data set 
unless the X-ray source is located outside the FOV, 
near a chip gap, or in a region of low exposure (such as a bad column) 
in one or more of the
instruments. In these cases we consider the measurement of the individual 
detector(s) more reliable, and make use of those instruments that do not
show any of the above mentioned problems at the respective source position. 
Table~\ref{tab:x-sources} summarizes all 58 detections and 
their characteristic X-ray parameters. The broad band counts have been extracted
from a circular region of radius $15^{\prime\prime}$ centered on the source. 
The background counts were obtained from the background map. 
After the background was subtracted the source counts were corrected for the
flux in the PSF wings outside the extraction radius ($\sim 30$\,\%). 
The exposure times given in Table~\ref{tab:x-sources} were obtained from the
broad band exposure map. Hardness ratios are defined and discussed in 
Sect.~\ref{sect:spec}. 
The merged EPIC pn and MOS image is shown in Fig.~\ref{fig:mergedimage},
and the position of all detected sources is marked. 

\begin{table*}\scriptsize
\begin{center}
\caption{X-ray sources detected in the EPIC observation of Cha\,I at $ML > 20$. The columns denote
X-ray source number, instrument by which the source was detected, X-ray position, maximum likelihood of source existence, number of broad band source counts, exposure time, hardness ratios, optical/IR counterpart, and distance between the X-ray source and the counterpart. For sources detected in the merged image we list the hardness ratios computed from the corresponding source found in pn, and no hardness ratio is computed if the source is detected only in the merged image.}
\label{tab:x-sources}
\newcolumntype{d}[1]{D{.}{.}{#1}}
\begin{tabular}{llrrrr@{\hspace{-0.03cm}}c@{\hspace{-0.04cm}}rrrrlr} \\ \hline
Designation & Instr & $\alpha_{\rm x,2000}$ & $\delta_{\rm x,2000}$ & $ML$ & \multicolumn{3}{c}{Cts} & \multicolumn{1}{c}{Exp} & \multicolumn{1}{c}{HR\,1} & \multicolumn{1}{c}{HR\,2} & Identification & $\Delta_{\rm ox}$ \\ 
            &       &                       &                       &      & & &                     & \multicolumn{1}{c}{[s]} &                           &                           &                & [$^{\prime\prime}$] \\ \hline
\hline
XMM-Cha\,I-1  & ALL  & 11 04 32.7 & -77 28 49.0 &    143.5 &  108.6 & $ \pm $ &   10.4 &  31309.5 & $   0.76 \pm  0.11$ & $  -0.02 \pm  0.20$ & & \\
XMM-Cha\,I-2  & ALL  & 11 04 42.4 & -77 41 55.7 &    320.4 &  166.8 & $ \pm $ &   12.9 &  42316.2 & $< -0.16$           & $< -0.85$           & BYB\,18   & 2.2 \\
XMM-Cha\,I-3  & M2   & 11 05 03.6 & -77 39 44.6 &     46.7 &   33.8 & $ \pm $ &    5.8 &  49723.4 & $>  0.84$           & $   0.16 \pm  0.24$ & 2M\,110506-773944 & 1.4 \\
XMM-Cha\,I-4  & ALL  & 11 05 07.0 & -77 33 42.9 &     21.3 &   37.4 & $ \pm $ &    6.1 &  47340.8 &  $-$                 &  $-$                 & 2M\,110507-773338 & 5.0 \\
XMM-Cha\,I-5  & ALL  & 11 05 09.3 & -77 43 28.9 &     22.0 &   48.0 & $ \pm $ &    6.9 &  44590.1 & $>  0.82$           & $   0.01 \pm  0.26$ & & \\
XMM-Cha\,I-6  & ALL  & 11 05 43.0 & -77 26 50.6 &   1642.2 &  508.1 & $ \pm $ &   22.5 &  34147.7 & $  -0.28 \pm  0.08$ & $  -0.93 \pm  0.04$ & CHXR-15   & 1.9 \\
XMM-Cha\,I-7  & ALL  & 11 06 15.3 & -77 37 50.3 &    265.8 &  194.4 & $ \pm $ &   13.9 &  71956.2 & $   0.45 \pm  0.13$ & $  -0.65 \pm  0.12$ & CCE\,98-19  & 2.2 \\
XMM-Cha\,I-8  & ALL  & 11 06 15.7 & -77 21 56.1 &  11306.0 & 2337.6 & $ \pm $ &   48.3 &  14244.0 & $   0.47 \pm  0.03$ & $  -0.58 \pm  0.03$ & Ced\,110-IRS2& 1.0 \\
XMM-Cha\,I-9  & ALL  & 11 06 28.8 & -77 37 33.0 &    600.9 &  342.6 & $ \pm $ &   18.5 &  75135.8 & $   0.34 \pm  0.10$ & $  -0.71 \pm  0.09$ & CHXR-73   & 0.3 \\
XMM-Cha\,I-10 & M2   & 11 06 44.5 & -77 26 32.5 &     61.4 &   46.6 & $ \pm $ &    6.8 &  38064.8 & $   0.43 \pm  0.18$ & $  -0.94 \pm  0.07$ & UX\,Cha & 3.3 \\
XMM-Cha\,I-11 & M2   & 11 06 45.1 & -77 27 00.7 &    562.5 &  180.7 & $ \pm $ &   13.4 &  17263.0 & $   0.70 \pm  0.07$ & $  -0.69 \pm  0.08$ & CHXR-20   & 2.8 \\
XMM-Cha\,I-12 & ALL  & 11 06 57.6 & -77 42 10.1 &   3034.5 & 1037.6 & $ \pm $ &   32.2 &  68823.7 & $   0.11 \pm  0.06$ & $  -0.80 \pm  0.04$ & CHXR-74   & 0.8 \\
XMM-Cha\,I-13 & ALL  & 11 07 11.4 & -77 46 37.2 &   2151.1 &  778.4 & $ \pm $ &   27.9 &  47614.2 & $   0.39 \pm  0.07$ & $  -0.64 \pm  0.06$ & CHXR-21    & 4.1 \\
XMM-Cha\,I-14 & M1   & 11 07 12.1 & -77 38 57.8 &     34.2 &   33.4 & $ \pm $ &    5.8 &  38792.2 & $>  0.84$           & $   0.33 \pm  0.23$ & & \\
XMM-Cha\,I-15 & M2   & 11 07 12.8 & -77 43 48.2 &    310.7 &  151.4 & $ \pm $ &   12.3 &  53775.0 & $   0.39 \pm  0.10$ & $  -0.61 \pm  0.10$ & CHXR-22E/W & 2.8/7.6 \\
XMM-Cha\,I-16 & PN   & 11 07 13.0 & -77 23 06.9 &     34.9 &   35.8 & $ \pm $ &    6.0 &  21981.5 & $>  0.85$           & $   0.08 \pm  0.24$ & CCE\,98-23 & 11.6 \\
XMM-Cha\,I-17 & ALL  & 11 07 16.9 & -77 35 52.3 &     66.2 &   79.8 & $ \pm $ &    8.9 &  84142.5 & $  -0.23 \pm  0.18$ & $  -0.87 \pm  0.13$ & ChaH$\alpha$1    & 1.8 \\
XMM-Cha\,I-18 & ALL  & 11 07 19.5 & -77 36 38.6 &    211.4 &  181.9 & $ \pm $ &   13.5 &  85867.4 & $   0.95 \pm  0.04$ & $   0.46 \pm  0.13$ & & \\
XMM-Cha\,I-19 & PN   & 11 07 20.6 & -77 29 40.0 &    248.0 &  134.1 & $ \pm $ &   11.6 &  39992.5 & $  -0.61 \pm  0.09$ & $  -0.87 \pm  0.12$ & KG\,2001-5 & 4.0 \\
XMM-Cha\,I-20 & M2   & 11 07 20.7 & -77 38 07.1 &   6182.4 & 1391.6 & $ \pm $ &   37.3 &  66026.2 & $   0.31 \pm  0.04$ & $  -0.78 \pm  0.03$ & LH$\alpha$332-17 & 3.0 \\
XMM-Cha\,I-21 & M1   & 11 07 32.9 & -77 28 25.9 &   1289.1 &  362.1 & $ \pm $ &   19.0 &  45877.1 & $  -0.13 \pm  0.07$ & $  -0.81 \pm  0.06$ & CHXR-25   & 2.1 \\
XMM-Cha\,I-22 & ALL  & 11 07 35.1 & -77 34 49.1 &   1997.0 &  849.7 & $ \pm $ &   29.1 &  82940.4 & $   0.13 \pm  0.07$ & $  -0.90 \pm  0.03$ & CHXR-76   & 2.0 \\
XMM-Cha\,I-23 & ALL  & 11 07 37.1 & -77 33 33.2 &   6652.4 & 1980.5 & $ \pm $ &   44.5 &  77381.9 & $   0.62 \pm  0.03$ & $  -0.58 \pm  0.04$ & CHXR-26   & 0.4 \\
XMM-Cha\,I-24 & ALL  & 11 07 37.5 & -77 47 14.3 &    107.7 &   94.7 & $ \pm $ &    9.7 &  46535.7 & $  -0.09 \pm  0.20$ & $  -0.76 \pm  0.18$ & KG\,2001-19  & 5.5 \\
XMM-Cha\,I-25 & ALL  & 11 07 37.8 & -77 35 29.1 &     68.9 &   83.1 & $ \pm $ &    9.1 &  85628.3 & $< -0.33$           & $< -0.64$           & ChaH$\alpha$7    & 2.2 \\
XMM-Cha\,I-26 & ALL  & 11 07 43.1 & -77 33 57.9 &     60.5 &  106.4 & $ \pm $ &   10.3 &  79734.8 & $   0.04 \pm  0.19$ & $  -0.85 \pm  0.13$ & ChaH$\alpha$2    & 2.4 \\
XMM-Cha\,I-27 & M2   & 11 07 43.3 & -77 39 39.6 &    476.3 &  233.4 & $ \pm $ &   15.3 &  37801.9 & $   0.67 \pm  0.06$ & $  -0.59 \pm  0.08$ & HM\,15      & 5.3 \\
XMM-Cha\,I-28 & ALL  & 11 07 46.1 & -77 40 04.6 &     24.3 &   84.4 & $ \pm $ &    9.2 &  81971.4 &  $-$                 &  $-$                 & ChaH$\alpha$8    & 6.5 \\
XMM-Cha\,I-29 & ALL  & 11 07 52.2 & -77 36 57.4 &    279.0 &  229.3 & $ \pm $ &   15.1 &  89429.2 & $   0.10 \pm  0.14$ & $  -1.00 \pm  0.01$ & ChaH$\alpha$3    & 3.5 \\
XMM-Cha\,I-30 & ALL  & 11 07 55.9 & -77 27 24.5 &  12386.6 & 2842.1 & $ \pm $ &   53.3 &  44689.5 & $   0.34 \pm  0.03$ & $  -0.72 \pm  0.03$ & CHX-10a   & 3.6 \\
XMM-Cha\,I-31 & M2   & 11 07 57.5 & -77 38 44.7 &    238.0 &  260.3 & $ \pm $ &   16.1 &  86223.8 & $   0.84 \pm  0.04$ & $   0.15 \pm  0.09$ & HM\,16      & 5.3 \\
XMM-Cha\,I-32 & ALL  & 11 08 01.5 & -77 42 27.8 &  38290.7 & 8011.2 & $ \pm $ &   89.5 &  71190.8 & $   0.17 \pm  0.02$ & $  -0.68 \pm  0.02$ & VW\,Cha      & 1.6 \\
XMM-Cha\,I-33 & M2   & 11 08 02.6 & -77 38 42.1 &    968.0 &  330.6 & $ \pm $ &   18.2 &  86564.1 & $   0.95 \pm  0.02$ & $   0.47 \pm  0.07$ & CCE\,98-32  & 5.4 \\
XMM-Cha\,I-34 & ALL  & 11 08 03.7 & -77 39 13.1 &    116.0 &  171.4 & $ \pm $ &   13.1 &  84836.8 & $   0.68 \pm  0.10$ & $  -0.28 \pm  0.15$ & HD\,97048   & 6.1 \\
XMM-Cha\,I-35 & ALL  & 11 08 11.2 & -77 48 15.5 &   4170.3 & 1202.0 & $ \pm $ &   34.7 &  41846.1 & $   0.92 \pm  0.02$ & $   0.05 \pm  0.06$ & WGA\,J1108.1-7748 & 1.6 \\
XMM-Cha\,I-36 & M1   & 11 08 13.0 & -77 34 20.4 &     59.6 &   70.5 & $ \pm $ &    8.4 &  81261.2 & $   0.51 \pm  0.14$ & $  -0.46 \pm  0.17$ & 2M\,110813-773410 & 9.8 \\
XMM-Cha\,I-37 & M2   & 11 08 14.7 & -77 33 52.1 &   4941.1 & 1111.5 & $ \pm $ &   33.3 &  39850.8 & $   0.35 \pm  0.04$ & $  -0.49 \pm  0.04$ & Glass\,I    & 5.6 \\
XMM-Cha\,I-38 & ALL  & 11 08 15.8 & -77 44 36.3 &    193.9 &  147.8 & $ \pm $ &   12.2 &  55259.6 & $  -0.26 \pm  0.16$ & $  -0.40 \pm  0.25$ & HM\,19       & 3.5 \\
XMM-Cha\,I-39 & ALL  & 11 08 16.4 & -77 44 09.9 &     50.0 &   63.0 & $ \pm $ &    7.9 &  60469.5 & $  -0.42 \pm  0.18$ & $  -0.30 \pm  0.36$ & ChaH$\alpha$13    & 3.4 \\
XMM-Cha\,I-40 & ALL  & 11 08 19.3 & -77 39 15.7 &    110.5 &  103.2 & $ \pm $ &   10.2 &  83166.5 & $  -0.05 \pm  0.19$ & $  -0.95 \pm  0.07$ & ChaH$\alpha$4    & 5.3 \\
XMM-Cha\,I-41 & ALL  & 11 08 23.9 & -77 41 45.3 &     80.1 &  109.2 & $ \pm $ &   10.5 &  72702.2 & $< -0.30$           & $< -0.65$           & ChaH$\alpha$5    & 3.1 \\
XMM-Cha\,I-42 & PN   & 11 08 26.2 & -77 34 20.6 &     20.7 &   51.1 & $ \pm $ &    7.1 &  79942.4 & $   0.52 \pm  0.16$ & $  -0.20 \pm  0.22$ & 2M\,110824-773422 & 5.7 \\
XMM-Cha\,I-43 & ALL  & 11 08 39.3 & -77 34 18.0 &     40.0 &   70.8 & $ \pm $ &    8.4 &  77325.5 & $< -0.42$           & $< -0.48$           & ChaH$\alpha$6    & 1.3 \\
XMM-Cha\,I-44 & ALL  & 11 08 45.1 & -77 33 02.7 &    242.8 &  196.4 & $ \pm $ &   14.0 &  70876.9 & $   0.88 \pm  0.06$ & $  -0.22 \pm  0.14$ & & \\
XMM-Cha\,I-45 & PN   & 11 08 54.7 & -77 32 09.6 &    120.9 &   93.0 & $ \pm $ &    9.6 &  59275.0 & $  -0.14 \pm  0.14$ & $  -0.83 \pm  0.11$ & CHXR-78  & 2.4 \\
XMM-Cha\,I-46 & ALL  & 11 09 07.1 & -77 40 03.2 &     75.6 &   99.3 & $ \pm $ &   10.0 &  71811.8 & $   0.89 \pm  0.08$ & $   0.35 \pm  0.19$ & & \\
XMM-Cha\,I-47 & M1   & 11 09 12.1 & -77 29 11.3 &   6138.8 & 1163.0 & $ \pm $ &   34.1 &  48242.7 & $  -0.05 \pm  0.04$ & $  -0.79 \pm  0.03$ & Sz\,30      & 1.0 \\
XMM-Cha\,I-48 & ALL  & 11 09 18.4 & -77 47 37.9 &   1038.1 &  372.8 & $ \pm $ &   19.3 &  33296.5 & $< -0.81$           & $< -0.73$           & KG\,2001-78  & 2.1 \\
XMM-Cha\,I-49 & ALL  & 11 09 20.1 & -77 31 22.3 &     21.8 &   53.8 & $ \pm $ &    7.3 &  57095.8 & $   0.73 \pm  0.20$ & $   0.73 \pm  0.21$ & & \\
XMM-Cha\,I-50 & ALL  & 11 09 27.1 & -77 25 29.0 &     20.0 &   32.3 & $ \pm $ &    5.7 &  31772.6 &  $-$                 &  $-$                 & & \\
XMM-Cha\,I-51 & ALL  & 11 09 45.5 & -77 40 33.5 &    121.9 &  110.7 & $ \pm $ &   10.5 &  60538.7 & $< -0.09$           & $< -0.84$           & KG\,2001-96 & 1.9 \\
XMM-Cha\,I-52 & ALL  & 11 09 47.9 & -77 26 28.0 &     33.2 &   54.6 & $ \pm $ &    7.4 &  34030.5 & $>  0.74$           & $  -0.31 \pm  0.30$ & KG\,2001-101 = BYB\,43 & 2.0 \\
XMM-Cha\,I-53 & ALL  & 11 09 57.6 & -77 40 42.5 &     24.3 &   43.0 & $ \pm $ &    6.6 &  56802.0 &  $-$                 &  $-$                 & 2M\,110955-774045 & 7.6 \\
XMM-Cha\,I-54 & PN   & 11 09 58.9 & -77 37 07.3 &   2343.6 & 1214.7 & $ \pm $ &   34.9 &  57935.2 & $   0.32 \pm  0.04$ & $  -0.57 \pm  0.04$ & WX\,Cha    & 1.6 \\
XMM-Cha\,I-55 & ALL  & 11 10 38.0 & -77 32 39.2 &   2074.4 &  723.5 & $ \pm $ &   26.9 &  44320.3 &  $-$                 &  $-$                 & CHXR-47   & 4.9 \\
XMM-Cha\,I-56 & ALL  & 11 10 51.5 & -77 43 10.9 &     80.6 &   81.3 & $ \pm $ &    9.0 &  40649.2 & $   0.75 \pm  0.14$ & $  -0.31 \pm  0.24$ & & \\
XMM-Cha\,I-57 & PN   & 11 10 55.0 & -77 36 15.2 &     20.2 &   27.9 & $ \pm $ &    5.3 &  45368.1 & $< -0.62$           & $> -0.03$           & & \\
XMM-Cha\,I-58 & M2   & 11 11 25.2 & -77 41 50.0 &     21.4 &   22.7 & $ \pm $ &    4.8 &  13717.4 & $   0.50 \pm  0.25$ & $  -0.72 \pm  0.22$ & & \\
\hline
\multicolumn{13}{l}{BYB - \protect\citey{Baud84.1}, 2\,M - 2\,MASS catalogue, CHXR - \protect\citey{Feigelson93.1}, CCE\,98 - \protect\citey{Cambresy98.1}, ChaH$\alpha$ - \protect\citey{Comeron00.1}, KG\,2001 - } \\ 
\multicolumn{13}{l}{\protect\citey{Kenyon01.1}, HM - \protect\citey{Henize73.1}, CHX - \protect\citey{Feigelson89.1}, Sz - \protect\citey{Schwartz77.1}.} \\
\end{tabular}
\end{center}
\end{table*}

\section{Nature of the X-ray Sources}\label{sect:nature}

\subsection{Identification of X-ray Sources}\label{subsect:identification}

In the past membership to the Cha\,I association has been assigned 
based on various indicators of youth, such as H$\alpha$ or X-ray emission, 
or Li absorption. 
We searched for counterparts to the {\em XMM-Newton} X-ray sources 
in the lists of stellar and substellar Cha\,I members  
published by \citey{Schwartz77.1}, \citey{Gauvin92.1},
\citey{Hartigan93.1}, \citey{Huenemoerder94.1}, LFH96, 
and CNK00, which in addition to presenting
new members also summarize other previous works.  
We also checked for X-ray 
sources at the position of NIR candidate members suggested by  
\citey{Cambresy98.1}, \citey{Comeron99.1}, \citey{Kenyon01.1}, 
\citey{Gomez01.1} and \citey{Carpenter02.1}, 
the mid-IR ISOCAM sources listed by \citey{Persi00.1},
and the Cha\,I candidates from a multi-band optical survey by 
\citey{LopezMarti04.1}. 
We mention in passing that further faint IR candidates 
have been published in regions of Cha\,I outside the FOV of our 
{\em XMM-Newton} pointing (e.g. \cite{Oasa99.1}), which are not of relevance for this study. 

For the identification of optical and IR counterparts to the X-ray 
sources we allowed for a maximum offset of $10^{\prime\prime}$. 
In total $42$ of the X-ray sources 
have published optical or IR counterparts. 
Most of the brighter X-ray sources can clearly be identified with known 
members of the Cha\,I association. 
Nearly all spectroscopically confirmed TTS members of Cha\,I in the FOV, 
and 9 of 13 VLM H$\alpha$ objects are detected. 
Evolutionary models place the dividing line between stars and brown dwarfs at 
spectral type M6.5. 
However, due to uncertainties in spectral classification, spectral type
to temperature conversion and the models themselves, CNK00 have suggested 
M7.5 as a conservative location of the substellar boundary in Cha\,I. Following this approach
the group of VLM H$\alpha$ objects in Cha\,I  
comprises 4 bona fide brown dwarfs, 
6 transition objects, 
and 3 VLM stars. 
The remaining counterparts to the X-ray sources include some of the recently identified 
faint NIR candidate members of Cha\,I. 
These objects are here shown to be X-ray emitters for the first time.
To learn more about the nature of the unidentified 
sources we searched the 2\,MASS archive and found counterparts to an additional five 
{\em XMM-Newton} sources. 
Finally, a search in SIMBAD\footnote{The SIMBAD data base is operated by the CDS and can 
be accessed via the URL http://simbad.u-strasbg.fr/sim-fid.pl} 
turned up a BL\,Lac object as counterpart to XMM-Cha\,I-35. 

A cross-check by visual inspection showed that one probable counterpart was
missed by the automatic identification procedure. This source (XMM-Cha\,I-16) 
is at the edge of the field, i.e. at large off-axis angle where the point spread
function is enlarged. The X-ray source is at a distance of $11.6^{\prime\prime}$ from the
cloud member CCE\,98-23, detected first in NIR photometry (\cite{Cambresy98.1})
and confirmed with a NIR spectrum (\cite{Gomez03.1}). 
We add this identification to the list in Table~\ref{tab:x-sources}. 

A list with optical/IR data and stellar parameters relevant for the interpretation of 
the X-ray emission is compiled in Table~\ref{tab:opt_lx} for all known and potential 
Cha\,I members in the {\em XMM-Newton} FOV. 
In the following we describe how these parameters were derived. 
\begin{table*}\scriptsize 
\begin{center}
\caption{Optical and IR properties for all Cha\,I members and candidate members in the {\em XMM-Newton} FOV. The H$\alpha$ equivalent width ($W_{\rm H\alpha}$) is negative for emission. If more than one H$\alpha$ measurement is available for a given star we list the maximum and the minimum value observed. The T Tauri type given in column~6 of the same table is assigned on basis of $W_{\rm H\alpha}$ using the criteria put forth by \protect\citey{Martin98.1}. Columns~7 and~8 represent the extinction and magnitude used to derive $L_{\rm bol}$; referring to the band given in column~9. The bolometric luminosity, effective temperature, age and mass have been computed as described in Sect.~\ref{subsect:lbol_age}. $L_{\rm bol}$ has been calculated for $d=160$\,pc. We assign an age of $5 \times 10^5$\,yrs to objects on the birthline.}
\label{tab:opt_lx}
\begin{tabular}{llrrrcrrrrrrrr} \hline
Designation  &  \myrule SpT & Ref. & $W_{\rm H\alpha}$ & Ref & TTS & $A_{\rm J/I}$ & $J/I$   & Band & Ref. & $\log{(L_{\rm bol}/L_\odot)}$  & $\log{T_{\rm eff}}$ & Age   & Mass              \\
             &  \myrule     &      & [\AA]             &     &     & [mag]         &   [mag] &      &      &                               & [K]                 & [Myr] & [${\rm M_\odot}$] \\
\hline
\multicolumn{14}{c}{\myrule Spectroscopically confirmed Cha\,I members} \\
\hline
KG\,2001-19    & M5      & (1) & $-7.3$           & (1)     & W & $0.3 $ & $12.24$ & J & (1)     & $-1.26$ &  $3.51$ &      4.0 & 0.18 \\
CHXR-21        & K7      & (2) & $-8.0...+1.0$    & (7)/(2) & C & $1.0 $ & $11.11$ & J & (5)     & $-0.36$ &  $3.61$ &      6.0 & 0.8  \\
HM\,19         & M3.5    & (3) & $-5.0...-3.0$    & (3)/(7) & W & $0.8 $ & $11.29$ & J & (5)     & $-0.63$ &  $3.54$ &      2.0 & 0.3  \\
VW\,Cha        & M0.5    & (4) & $-146.9...-60.0$ & (3)/(7) & C & $0.8 $ & $ 8.63$ & J & (5)     & $+0.52$ &  $3.59$ &  $<$ 1.0 & 0.6  \\
CHXR-74        & M4.5    & (4) & $-8.3...-13.0$   & (8)/(4) & C & $0.9 $ & $11.55$ & J & (5)     & $-0.72$ &  $3.53$ &      1.5 & 0.25 \\
BYB\,18        & M4      & (5) & $-$              & $-$     & W & $0.5 $ & $11.80$ & J & (5)     & $-0.98$ &  $3.53$ &      3.0 & 0.25 \\
KG\,2001-96    & M6.5    & (1) & $-$              & $-$     & W & $0.5 $ & $12.36$ & J & (1)     & $-1.24$ &  $3.48$ &$\leq$2.0 & 0.11 \\
HD\,97048      & A0      & (3) & $-30.0...-37.0$  & (3)/(7) &$-$& $0.4*$ & $ 7.41$ & J & (3)     & $+1.58$ &  $4.02$ &      2.0 & 2.5  \\
CCE\,98-19     & M2.5    & (5) & $-$              & $-$     & W & $2.9 $ & $12.59$ & J & (5)     & $-0.25$ &  $3.55$ &      1.0 & 0.35 \\
CHXR-73        & M4.5    & (4) & $-$              & $-$     & W & $1.8 $ & $12.60$ & J & (5)     & $-0.78$ &  $3.53$ &      2.0 & 0.25 \\
CHXR-76=BYB\,34& M5      & (4) & $-5.5...-7.2$    & (4)/(8) & W & $0.8 $ & $12.19$ & J & (5)     & $-1.04$ &  $3.51$ &      2.5 & 0.18 \\
CHXR-26        & K7      & (2) & $-$              & $-$     & W & $2.3 $ & $11.41$ & J & (5)     & $+0.04$ &  $3.61$ &      2.0 & 0.8  \\
CHXR-47        & K3      & (2) & $-0.4$           & (9)     & W & $1.4 $ & $ 8.45$ & J & (2)     & $+0.93$ &  $3.67$ &  $<$ 0.5 & 1.5  \\
CHX-10a        & M1      & (2) & $<-0.2$          & (10)    & W & $0.8 $ & $ 8.37$ & J & (2)     & $+0.61$ &  $3.57$ &  $<$ 0.5 & 0.5  \\
CHXR-15        & M5      & (2) & $-8.0$           & (9)     & W & $0.1 $ & $11.21$ & J & (2)     & $-0.93$ &  $3.51$ &      2.0 & 0.18 \\
KG\,2001-101   & M1      & (1) & $-178$           & (1)     & C & $2.1 $ & $12.77$ & J & (1)     & $-0.63$ &  $3.57$ &      4.0 & 0.5  \\
Ced110-IRS2    & G2      & (3) & $<-1.2$          & (3)     & W & $1.0*$ & $ 7.64$ & J & (3)     & $+1.26$ &  $3.77$ &      1.0 & 2.25 \\
WX\,Cha        & K7      & (3) & $-65.5...-90.0$  & (3)/(7) & C & $0.6*$ & $10.00$ & J & (3)     & $-0.07$ &  $3.61$ &      2.5 & 0.8  \\
CHXR-78\,C     & M5.5    & (4) & $-3.2...-13.4$   & (4)/(8) & C & $0.6 $ & $12.36$ & J & (5)     & $-1.19$ &  $3.51$ &      3.0 & 0.18 \\
CCE\,98-23     & M1      & (5) & $-$              &         & C & $8.6 $ & $17.55$ & J & (5),(11) & $+0.06$ &  $3.57$ &  $<$ 1.0 & 0.45 \\
Sz\,30         & M0      & (3) & $-5.2$           & (3)     & W & $0.3*$ & $ 9.70$ & J & (3)/(12) & $-0.11$ &  $3.59$ &      1.5 & 0.6  \\
CHXR-25        & M4      & (2) & $1.2$            & (2)     & W & $0.0 $ & $11.69$ & J & (2)     & $-1.14$ &  $3.53$ &      5.0 & 0.25 \\
CHXR-22\,E     & M3      & (2) & $-$              & $-$     & W & $1.8 $ & $10.18$ & J & (2)     & $+0.22$ &  $3.54$ &      0.5 & 0.3  \\
CHXR-22\,W     & K7      & (2) & $-$              & $-$     & W & $1.9 $ & $12.56$ & J & (2)     & $-0.58$ &  $3.61$ &     10.0 & 0.8  \\
HM\,15         & M1      & (4) & $-47.9...-70.0$  & (3)/(4) & C & $1.3 $ & $10.22$ & J & (5)     & $+0.07$ &  $3.57$ &  $<$ 1.0 & 0.5  \\
CCE\,98-32     & M2      & (5) & $-$              & $-$     & C & $3.4 $ & $11.51$ & J & (5)     & $+0.39$ &  $3.55$ &      0.5 & 0.35 \\
HM\,16         & K7      & (4) & $-61.0...-75.0$  & (4)/(7) & C & $2.3 $ & $ 9.71$ & J & (5)     & $+0.72$ &  $3.61$ &  $<$ 0.5 & 0.8  \\
LH$\alpha$332-17&G2      & (3) & $-17.2$          & (3)     & C & $1.2 $ & $ 7.99$ & J & (5)     & $+1.20$ &  $3.77$ &      1.5 & 2.4  \\
Glass\,I       & K7      & (4) & $-4.0...-11.0$   & (3)/(4) & W & $1.5 $ & $ 8.58$ & J & (5)     & $+0.86$ &  $3.61$ &  $<$ 0.5 & 0.8  \\
CHXR-20        & M0      & (2) & $0.1$            & (2)     & W & $0.6 $ & $ 9.88$ & J & (2)     & $-0.06$ &  $3.59$ &      1.0 & 0.6  \\
ChaH$\alpha$1  & M7.5    & (6) & $-34.5...-59.0$  & (8)/(4) & C & $0.11$ & $16.17$ & I & (6)     & $-1.95$ &  $3.46$ &      5.0 & 0.06 \\
ChaH$\alpha$2  & M6.5    & (6) & $-39.0...-71.0$  & (6)/(8) & C & $0.40$ & $15.08$ & I & (6)     & $-1.58$ &  $3.48$ &      4.0 & 0.09 \\
ChaH$\alpha$3  & M7      & (6) & $-4.5...-14.4$   & (6)/(8) & W & $0.16$ & $14.89$ & I & (6)     & $-1.45$ &  $3.46$ &$\leq$1.6 & 0.07 \\
ChaH$\alpha$4  & M6      & (6) & $-4.7...-9.7$    & (4)/(8) & W & $0.29$ & $14.34$ & I & (6)     & $-1.33$ &  $3.48$ &      2.5 & 0.10 \\
ChaH$\alpha$5  & M6      & (6) & $-7.6...-8.0$    & (4)/(8) & W & $0.47$ & $14.68$ & I & (6)     & $-1.40$ &  $3.48$ &      3.2 & 0.10 \\
ChaH$\alpha$6  & M7      & (6) & $-59.0...-61.7$  & (6)/(8) & C & $0.13$ & $15.13$ & I & (6)     & $-1.56$ &  $3.46$ &      1.6 & 0.06 \\
ChaH$\alpha$7  & M8      & (6) & $-35.0...-45.0$  & (8)/(6) & C & $0.14$ & $16.86$ & I & (6)     & $-2.18$ &  $3.44$ &      6.3 & 0.04 \\
ChaH$\alpha$8  & M6.5    & (6) & $-8.4...-9.0$    & (8)/(6) & W & $0.34$ & $15.47$ & I & (6)     & $-1.76$ &  $3.48$ &      5.0 & 0.10 \\
ChaH$\alpha$9  & M6      & (6) & $-16.0$          & (6)     & W & $0.76$ & $17.34$ & I & (6)     & $-2.34$ &  $3.48$ &     30.0 & 0.09 \\
ChaH$\alpha$10 & M7.5    & (6) & $-9.0$           & (6)     & W & $0.05$ & $16.90$ & I & (6)     & $-2.27$ &  $3.46$ &     12.0 & 0.06 \\
ChaH$\alpha$11 & M8      & (6) & $-23.0$          & (6)     & C & $0.0 $ & $17.35$ & I & (6)     & $-2.44$ &  $3.44$ &     12.0 & 0.04 \\
ChaH$\alpha$12 & M7      & (6) & $-20.0$          & (6)     & C & $0.5:$ & $15.58$ & I & (6)     & $-1.59$ &  $3.46$ &      1.6 & 0.06 \\
ChaH$\alpha$13 & M5      & (6) & $-11.0$          & (6)     & W & $0.89$ & $14.09$ & I & (6)     & $-1.06$ &  $3.51$ &      2.0 & 0.18 \\
IR\,Nebula     & M5      & (5) & $-$              & $-$     &   & $3.2 $ & $11.12$ & J & (5)     & $+0.34$ &  $3.51$ &  $<$ 0.5 & 0.18 \\
\hline
\multicolumn{14}{c}{\myrule Photometric Cha\,I candidates} \\
\hline
UX\,Cha        & M1.5    &     & $<-5.0$          & (7)     & W & $0.9 $ & $10.81$ & J & 2\,MASS & $-0.32$ &  $3.57$ &      2.0 & 0.4  \\
KG\,2001-78      & A2      &     & $-$              & $-$     &   & $0.8 $ & $ 7.73$ & J & 2\,MASS & $+1.53$ &  $3.95$ &      4.0 & 2.0  \\
KG\,2001-5       & M1      &     & $-$              & $-$     &   & $0.2 $ & $11.31$ & J & (12)    & $-0.80$ &  $3.57$ &      3.0 & 0.45 \\
2M\,110955-774045  & M1  &     & $-$              & $-$     &   & $1.5 $ & $15.90$ & J & 2\,MASS & $-2.12$ &  $3.57$ &          &      \\
2M\,110507-773338  & M3  &     & $-$              & $-$     &   & $1.2 $ & $15.96$ & J & 2\,MASS & $-2.34$ &  $3.54$ &          &      \\
2M\,110824-773422  & M0.5&     & $-$              & $-$     &   & $0.8 $ & $14.37$ & J & 2\,MASS & $-1.78$ &  $3.59$ &          &      \\
2M\,110813-773410  & M2  &     & $-$              & $-$     &   & $0.9 $ & $16.24$ & J & 2\,MASS & $-2.51$ &  $3.55$ &          &      \\
2M\,110506-773944  & M0  &     & $-$              & $-$     &   & $2.3 $ & $13.29$ & J & 2\,MASS & $-0.74$ &  $3.59$ &      8.0 & 0.6  \\
\hline
\multicolumn{14}{l}{$^*$ $A_{\rm J}$ calculated from the literature values for $A_{\rm V}$ using the RL85-extinction law.} \\
\multicolumn{14}{l}{(1) - \protect\citey{Comeron04.1}, (2) - \protect\citey{Lawson96.1}, (3) - \protect\citey{Gauvin92.1}, (4) - \protect\citey{Comeron99.1}, (5) - \protect\citey{Gomez03.1}, (6) - \protect\citey{Comeron00.1},} \\
\multicolumn{14}{l}{(7) - \protect\citey{Hartigan93.1}, (8) - \protect\citey{Hartigan93.1}, (9) - \protect\citey{Neuhaeuser99.1}, (10) - \citey{Walter92.1}, (11) - \protect\citey{Cambresy98.1}, (12) - \citey{Kenyon01.1}.} \\
\end{tabular}
\end{center}
\end{table*}

\subsection{Spectral Types and Extinction}\label{subsect:extinc}

For optically faint stars extinction is often measured in 
the IR, e.g. through $A_{\rm J}$. In the IR dust extinction is composed of 
both interstellar reddening and absorption by circumstellar material. 
Determination of the extinction requires knowledge of the spectral type 
(used to obtain the intrinsic colors from empirical relations), 
the observed colors, and the assumption of an extinction law. 
In the ideal case spectral types are directly accessible from optical or IR spectroscopy. 
We use the spectral types and extinction ($A_{\rm J}$, $A_{\rm I}$) from the compilation by \citey{Gauvin92.1}, 
and from the spectroscopic surveys by LFH96, \citey{Comeron99.1}, CNK00, 
\citey{Gomez03.1}, and \citey{Comeron04.1}. 

\begin{figure}
\begin{center}
\resizebox{9cm}{!}{\includegraphics{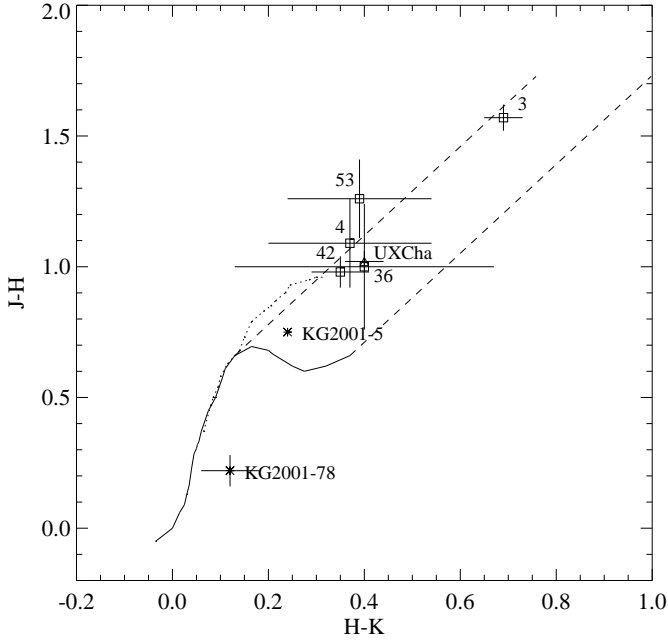}}
\caption{$J-H / H-K$ diagram showing the position of the eight X-ray detected Cha\,I 
candidates without spectroscopic information. The sources with 2\,MASS counterparts are
labeled with their X-ray source number, objects known before this study are referenced by
their previous designation. We use published NIR photometry 
for KG\,2001-5 (\protect\cite{Kenyon01.1}) and 2\,MASS data for all other objects. 
The solid and the dotted curves are the loci of dwarf and giant stars 
according to \protect\citey{Bessell88.1}. 
The dashed lines indicate a reddening corresponding to $A_{\rm V}=10$\,mag using 
the extinction law by RL85.}
\label{fig:jh_hk}
\end{center}
\end{figure}

For stars for which no spectroscopy is available,
the spectral types can be estimated from the photometry in color-color diagrams, 
projecting the objects back to the locus of unreddened dwarfs. The most widely used transformation
between spectral type and intrinsic colors is the one by \citey{Bessell88.1}.  
LFH96 have used this method for a number of X-ray emitting Cha\,I candidates discovered by 
{\em ROSAT}. The X-ray sources seen by {\em XMM-Newton} comprise 
some objects for which only visible and IR photometry is available so far: 
the highly extincted and poorly studied variable star 
UX\,Cha, two objects from the NIR survey by \citey{Kenyon01.1}, and five previously unknown
objects that we associate with 2\,MASS counterparts. 
In Fig.~\ref{fig:jh_hk} we show the $J-H / H-K$ color-color diagram for these objects. 
KG\,2001-78 lies bluewards of the reddening band according to the colors 
given by \citey{Kenyon01.1}. But according to 2\,MASS the object seems to be a moderately 
reddened A-type star. 
Since KG\,2001-78 is rather bright, we suspect that it was saturated in
one or more of the bands during the survey by \citey{Kenyon01.1}. 
Furtheron, we use its 2\,MASS photometry. 

All 8 objects in Fig.~\ref{fig:jh_hk} display colors of normal reddened photospheres, 
without indication for NIR excess from circumstellar material. 
Note that some of the newly identified 2\,MASS counterparts are faint in the NIR
($J \sim 16$\,mag; see Table~\ref{tab:opt_lx}), 
such that the error bars in Fig.~\ref{fig:jh_hk} extend over the full reddening band, 
making the assignment of spectral types practically impossible. 
We deredden the stars in the $J-H / H-K$ diagram along the 
extinction law of \citey{Rieke85.1} (hereafter RL85) back to the locus of main-sequence (MS) dwarfs 
(\cite{Bessell88.1}). 
In lack of spectroscopy the ambiguity by the two intersects of the reddening vector with the
MS locus can not be resolved. Based on the faintness of the objects we  
assume always the later spectral type, but keep in mind that 
this may lead to an underestimate of $A_{\rm V}$ by $1-3$\,mag. 
Recall also that any possible IR excess is not taken account of, 
and would result in an underestimate of the spectral type.

%
%

\subsection{Physical Parameters}\label{subsect:lbol_age}

%
%
\begin{figure}[t]
\begin{center}
\resizebox{9cm}{!}{\includegraphics{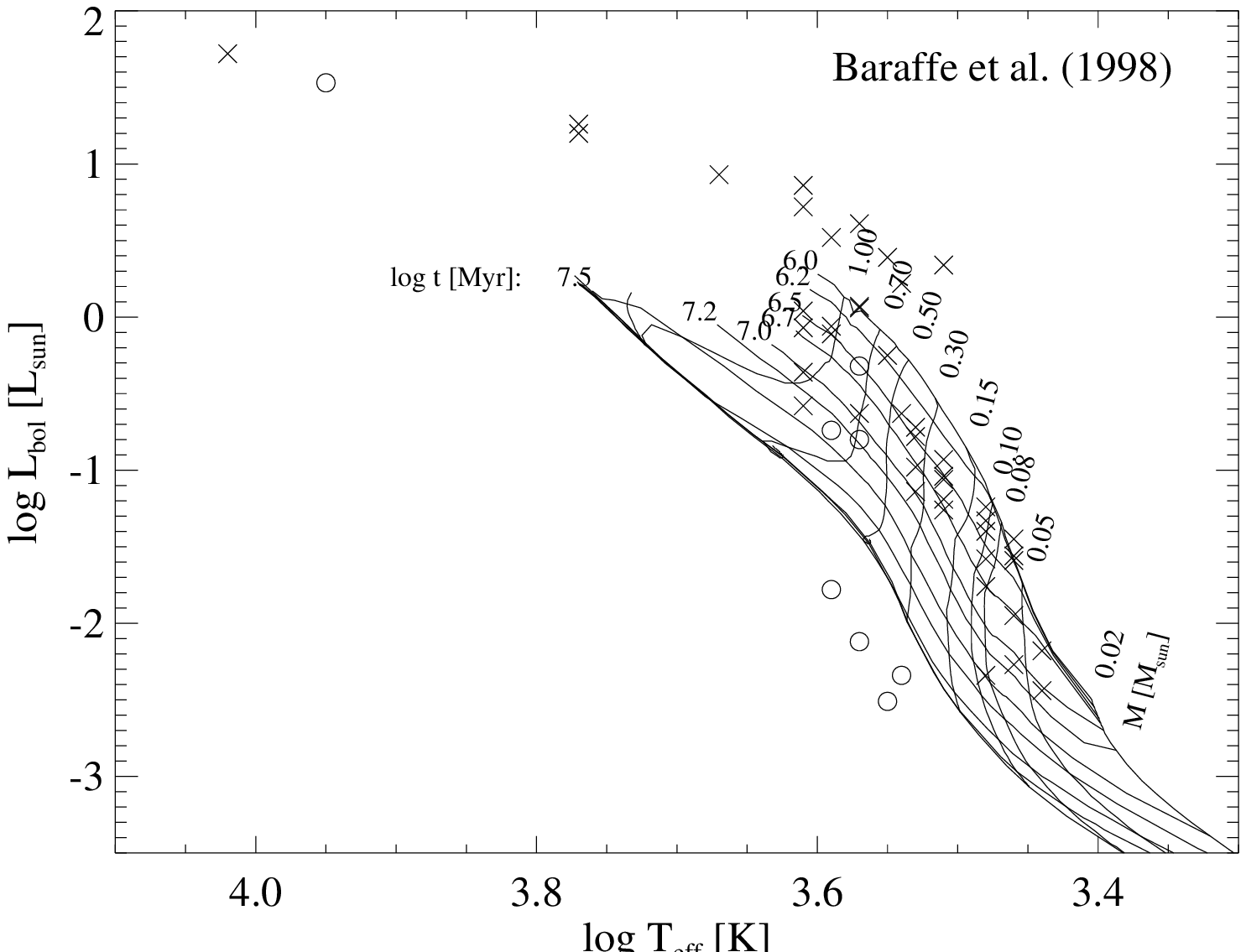}}
\resizebox{9cm}{!}{\includegraphics{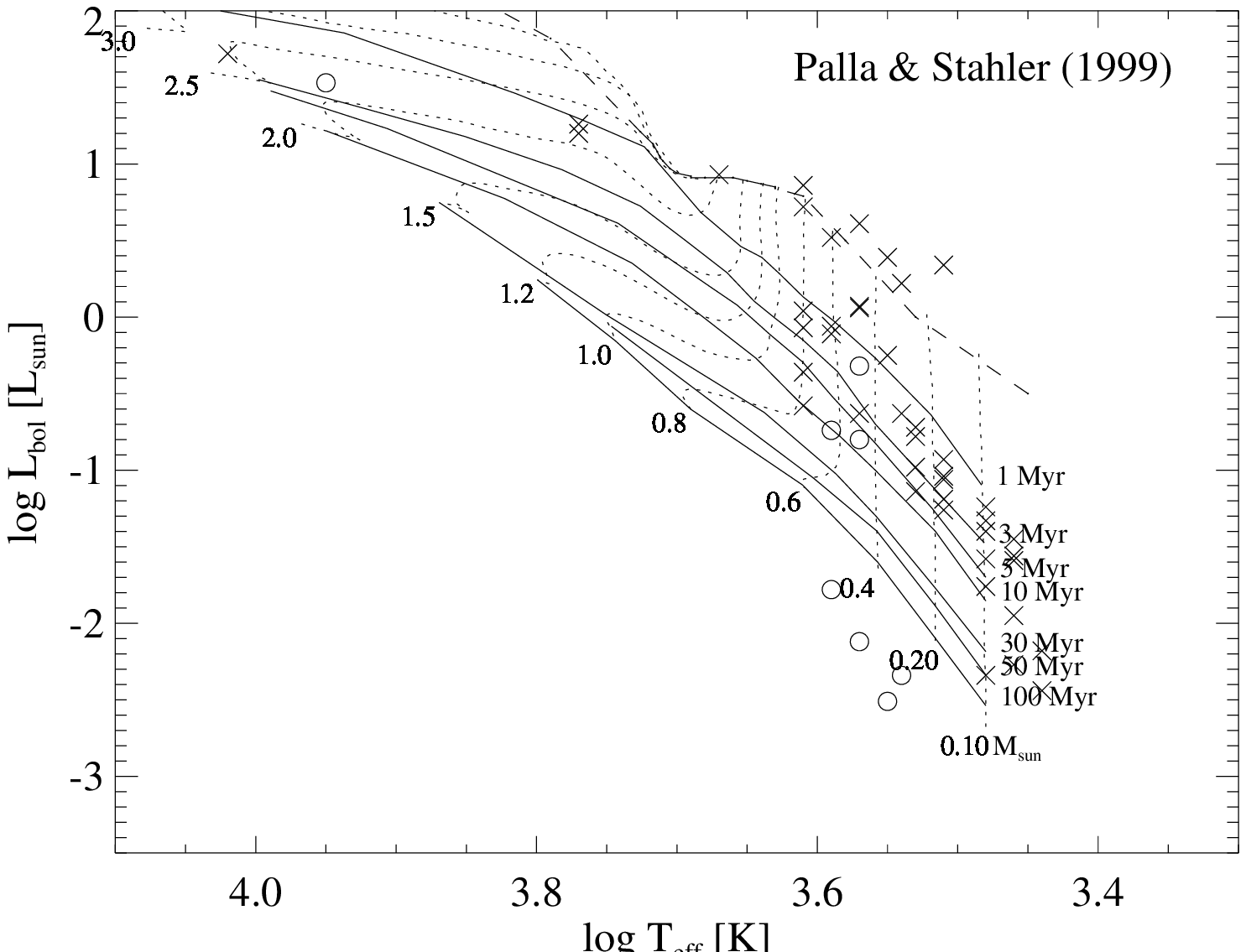}}
\caption{HRD for counterparts to X-ray sources in the {\em XMM-Newton} field. 
Tracks and isochrones that span the full range required to obtain masses and ages
for the Cha\,I sample are from \protect\citey{Baraffe98.1} (top) and 
\protect\citey{Palla99.1} (bottom). Counterparts of X-ray sources 
which are not spectroscopically studied are shown as circles, and spectroscopic cloud members
as x-points. The distance assumed is $160$\,pc.} 
\label{fig:hrd_tracks}
\end{center}
\end{figure}

In the last columns of Table~\ref{tab:opt_lx} we list the bolometric luminosity, 
effective temperature, age, and mass of the stars. 
Our estimate of the bolometric luminosity is based on the
$J$ band magnitude and extinction for stars with spectral type up to M6, because this band
is least affected by IR and UV excess. 
For objects later than M6 we use the $I$ band following CNK00's
adoption of the \citey{Zapatero97.1} empirical color-spectral type relationship. 
The bolometric luminosity is then given by 
\begin{displaymath}
\log{(L_{\rm bol}/L_\odot)} = 
\end{displaymath}
\begin{displaymath}
\hspace*{1cm} 1.86 - 0.4 \times [J + (V-J)_0 - A_{\rm J} - DM + BC_{\rm V}]
\label{eq:lbol}
\end{displaymath}
and an equivalent expression for the $I$ band. 
The distance modulus is $6.0$, corresponding to the distance of $160$\,pc
(\cite{Wichmann97.1}, \cite{Whittet97.1}) adopted throughout this paper. 
The intrinsic colors and bolometric corrections for dwarfs are
taken from \citey{Kenyon95.1}. Beyond spectral type M6 we use 
bolometric corrections $BC_{\rm I}$ as listed by CNK00, which represent a scale
intermediate between that of dwarfs and giants. 
Our luminosity determinations are in good agreement with those published in the literature.

The gravity of PMS stars is intermediate between that of dwarfs and
giants. Various temperature scales have been discussed by \citey{Luhman98.1}. 
We use the spectral type to temperature conversion of 
\citey{Kenyon95.1} for 
dwarfs. 
Beyond spectral type M6 we adopt the intermediate temperatures by \citey{Luhman99.1}. 

Masses and ages are derived in comparison to PMS evolutionary models. 
Although similar analyses have been performed before by various
authors we repeat this exercise in order to have a homogeneous sample that provides 
consistent {\em relative} ages and masses for all stars in the {\em XMM-Newton} field. 
The only set of PMS tracks that covers the temperature and luminosity range 
corresponding to the Cha\,I sample are the models by \citey{DAntona97.1}. 
However, these models overestimate the ages of the lowest mass stars. 
Therefore, we prefer a combination of the calculations by \citey{Baraffe98.1}
(with initial helium fraction $Y=0.275$ and initial mixing length parameter 
$\alpha_{\rm ML}=H_{\rm p}$) 
and by \citey{Palla99.1}. Fig.~\ref{fig:hrd_tracks} shows the {\em XMM-Newton} sample
of Cha\,I overplotted onto these models. 
The masses and ages from the \citey{Baraffe98.1} tracks and isochrones are adopted for 
objects with $M < 0.15\,M_\odot$, 
and those by \citey{Palla99.1} for all higher-mass stars. 
A detailed discussion of the masses and ages of the stars in the
present sample is not the scope of this paper. We will use the stellar parameters only to
examine correlations with X-ray emission.

The PMS models show that four of the five 2\,MASS counterparts to previously
unknown X-ray sources are below the zero-age MS assuming a distance of 160\,pc, 
indicating that they may not be related to the Chamaeleon cloud. 
Their $J$ band magnitudes are in the range or lower than those of the VLM members
of Cha\,I suggesting that these sources are located in the background. 
However, only a spectroscopic study will provide a conclusion on their nature.  
KG\,2001-78 and KG\,2001-5, and the 2\,MASS counterpart to 
X-ray source XMM-Cha\,I-3 are placed for the first time in the HRD. Their position is compatible
with them being PMS stars, i.e. they are candidate Cha\,I members.

\section{Spectral Analysis}\label{sect:spec}

For the spectral analysis source photons were extracted from a circular region 
centered on the positions given in Table~\ref{tab:x-sources}, 
and the background was extracted from an adjacent circular area on the same CCD chip. 
The source extraction radius was chosen individually such as to achieve high S/N 
without including contamination from neighboring sources. The radii that fullfill 
these conditions are between $15-30^{\prime\prime}$, with the exception of
XMM-Cha\,I-31 and XMM-Cha\,I-33 (see Sect.~\ref{subsect:hd97048_group}). 
Generally, the background extraction region was much larger than the source extraction 
area, and the background was scaled accordingly. 
A redistribution matrix and an ancilliary reponse file were calculated
for each source using the {\em XMM}-SAS. 
The spectra were background subtracted and binned to a minimum of 
$15$~counts per bin. For the brighter sources we binned up to $30$~counts per bin.  
Spectral fitting was carried out in the XSPEC environment, version 11.2.0. 

We performed a spectral analysis for all X-ray sources with 
$> 150$ counts in Table~\ref{tab:x-sources}.
This rather low threshold results in considerable
uncertainties of the spectral model for the faintest sources of the spectral
sample, as a consequence of low statistics. 
A detailed modelling of the temperature structure in the coronae of our targets
is out of reach with the presently available observations. Nevertheless, these 
{\em XMM-Newton} data provide for the first time an estimate for the 
temperature(s) dominating in the coronae of these stars, 
and they allow us to identify peculiar X-ray sources. 

In most cases we modeled only the EPIC pn spectrum because of its higher sensitivity
as compared to EPIC MOS.
The exception are those sources that are located on chip gaps in the pn. 
Due to an unfortunate roll-angle this applies to some of the brightest X-ray emitters 
in this field. 
For these latter sources we jointly modeled the MOS\,1 and MOS\,2 spectrum. 
CHXR-25 falls on a gap in both pn and MOS\,2, and we use only MOS\,1 for the spectral
analysis. 

In a first approach we tried to describe each X-ray spectrum with a one-temperature (1-T) 
thermal plasma model (MEKAL; \cite{Mewe85.1}, \cite{Mewe95.1}) plus photo-absorption 
term with the atomic cross-sections of \citey{Morrison83.1}. 
Subsolar global abundances from \citey{Anders89.1} 
were chosen motivated by previous results on PMS stars. 
We adapted the most widely used value of  
$Z = 0.3\,Z_\odot$ in order to enable direct comparison to previous {\em Chandra} and
{\em XMM-Newton} studies in other star forming regions (e.g. \cite{Nakajima03.1},
\cite{Feigelson02.1}, \cite{Preibisch03.1}).  
Our simplest model has three free parameters: X-ray temperature, emission
measure, and hydrogen column density of the absorbing gas. 

For some X-ray sources this 1-T model 
is obviously inadequate (either $\chi^2_{\rm red} > 2$ or systematic residuals in a
certain spectral range). Allowing the global abundance parameter to vary improves the
result in about half of the cases, however, with unreasonably low values for the
abundance ($Z \sim 0.05\,Z_\odot$), which we consider unphysical. 
Adding a second temperature component (2-T model) instead provides a statistically satisfying 
solution in most cases. 
Due to the low statistics and the low-energy cutoff of $0.3$\,keV, 
for most of the sources the $N_{\rm H}$ values are not well-constrained, 
and tend to be correlated with the emission measure. 

Since our targets are subject to considerable extinction, both from material in the
cloud and from circumstellar material, the absorbing column $N_{\rm H}$
is an important parameter. 
Comparison of the X-ray absorption to measurements of optical extinction
can be helpful to check the X-ray fitting results. 
All $A_{\rm J}$ values adopted or computed in this paper are listed in Table~\ref{tab:opt_lx}.  
We use the RL85 extinction law to obtain $A_{\rm V}$, and the relation by 
\citey{Paresce84.1} to convert $A_{\rm V}$ to hydrogen column density, $N_{\rm H, V}$. 
Comparing $N_{\rm H, V}$ to the values from the X-ray spectrum ($N_{\rm H, X}$) we find that
in about 2/3 of the cases the value expected from the 
standard galactic relation 
is bracketed by the 90\,\% error bars provided by the XSPEC fits. 
Discrepancies arise for some of the X-ray brightest sources, for which $N_{\rm H, X}$
is significantly higher or lower than $N_{\rm H, V}$. However, due to the low-number statistics
our data are not sufficient to draw conclusions on a possible deviation from the standard 
extinction law. 

To obtain as homogeneous a set of spectral fits as possible we 
applied a 2-T model to each source, and held the column density 
fixed on the value corresponding to the opt./IR value ($N_{\rm H,V}$). 
These fits also result in acceptable solutions ($\chi^2_{\rm red} \sim 1$), 
i.e. they can not be statistically distinguished from the fits with free column density. 
We consider these latter models more realistic, as motivated below.
The best fit parameters for all bright X-ray sources with a counterpart Cha\,I 
member or candidate member are summarized in Table~\ref{tab:specparams_nhlit}. 

Three objects from the spectral sample have no
counterpart among the known or suspected Cha\,I members, and the thermal model
results in a very high X-ray temperature.  
One of them (XMM-Cha\,I-35) is a BL\,Lac object, the other two have no known
optical counterpart in SIMBAD and in the literature. 
These three objects can be described equally well by a power-law model,
consistent with them being extragalactic 
(see discussion in Sect.~\ref{subsect:uiden}). 
\begin{table*}
\begin{center}
\caption{Spectral parameters of bright X-ray sources in the Cha\,I {\em XMM-Newton} field. The columns represent: source number, identification, number of counts in the background subtracted spectrum, reduced $\chi^2$ of bestfit, absorbing hydrogen column density derived from $A_{\rm J}$ using the relation by \protect\citey{Paresce84.1} and held fixed during the fitting process, X-ray temperature(s) of 2-component thermal MEKAL model, and emission measures of each temperature component assuming a distance of 160\,pc.}
\label{tab:specparams_nhlit}
\newcolumntype{d}[1]{D{.}{.}{#1}}
\begin{tabular}{llrlrrrrr} \hline
Source  & Ident.          & Net counts  & \myrule $\chi^2_{\rm red}$ (d.o.f.) & $N_{\rm H}$                & $kT_{\rm 1}$           & $kT_{\rm 2}$   & $\log{EM_{\rm 1}}$       & $\log{EM_{\rm 2}}$     \\
        &                 &             & \myrule                          & [${\rm 10^{22}\,cm^{-2}}$] & [keV]                  & [keV]          & [${\rm erg/cm^2/s}$]    & [${\rm erg/cm^2/s}$]  \\
\hline
XMM-Cha\,I-2   & \myrule BYB\,18      & 92   & 1.13 (6)  & $=0.31$ & $0.21$ & $1.01$  & $52.21$ & $52.10$ \\
XMM-Cha\,I-6   & \myrule CHXR-15      & 264  & 1.00 (16) & $=0.06$ & $0.34$ & $1.12$  & $51.89$ & $52.46$ \\
XMM-Cha\,I-7   & \myrule CCE\,98-19   & 77   & 1.36 (3)  & $=1.82$ & $0.22$ & $79.2$  & $54.24$ & $51.82$ \\
XMM-Cha\,I-8   & \myrule Ced\,110\,IRS-2&2454& 1.27 (72) & $=0.60$ & $0.66$ & $2.15$  & $53.80$ & $53.89$ \\
XMM-Cha\,I-9   & \myrule CHXR-73      & 235  & 1.29 (6)  & $=1.13$ & $0.22$ & $0.86$  & $53.52$ & $52.75$ \\
XMM-Cha\,I-11  & \myrule CHXR-20      & 132  & 1.43 (11) & $=0.38$ & $0.33$ & $1.72$  & $52.41$ & $52.87$ \\
XMM-Cha\,I-12  & \myrule CHXR-74      & 457  & 1.20 (22) & $=0.57$ & $0.25$ & $1.24$  & $53.31$ & $52.74$ \\
XMM-Cha\,I-13  & \myrule CHXR-21      & 107  & 1.26 (9)  & $=0.63$ & $0.11$ & $1.27$  & $54.72$ & $52.57$ \\
XMM-Cha\,I-15  & \myrule CHXR-22      & 125  & 1.21 (15) & $=1.20$ & $0.20$ & $1.39$  & $54.15$ & $52.71$ \\
XMM-Cha\,I-20  & \myrule LH$\alpha$332-17&1165&1.40 (70) & $=0.75$ & $0.27$ & $1.03$  & $54.11$ & $53.42$ \\
XMM-Cha\,I-21  & \myrule CHXR-25      & 369  & 0.94 (16) & $=0.00$ & $0.51$ & $1.79$  & $52.43$ & $52.61$ \\
XMM-Cha\,I-22  & \myrule CHXR-76      & 352  & 0.96 (19) & $=0.50$ & $0.54$ & $1.00$  & $52.54$ & $52.52$ \\
XMM-Cha\,I-23  & \myrule CHXR-26      & 401  & 0.57 (22) & $=1.45$ & $0.26$ & $1.76$  & $54.31$ & $52.94$ \\
XMM-Cha\,I-27  & \myrule HM\,15       & 166  & 1.27 (11) & $=0.82$ & $0.22$ & $1.72$  & $53.49$ & $52.76$ \\
XMM-Cha\,I-29  & \myrule ChaH$\alpha$3& 110  & 0.99 (9)  & $=0.35$ & $0.59$ & $1.16$  & $51.49$ & $51.97$ \\
XMM-Cha\,I-30  & \myrule CHX-10a      & 1327 & 0.88 (47) & $=0.47$ & $0.61$ & $1.42$  & $53.07$ & $53.38$ \\
XMM-Cha\,I-31  & \myrule HM\,16       & 99   & 1.19 (8)  & $=1.45$ & $0.18$ & $6.40$  & $54.12$ & $52.62$ \\
XMM-Cha\,I-32  & \myrule VW\,Cha      & 3328 & 1.29 (93) & $=0.50$ & $0.32$ & $1.57$  & $53.80$ & $53.58$ \\
XMM-Cha\,I-33  & \myrule CCE\,98-32   & 92   & 0.66 (7)  & $=2.14$ & $0.25$ & $17.93$ & $53.99$ & $52.82$ \\
XMM-Cha\,I-34  & \myrule HD\,97048    & 74   & 1.58 (6)  & $=0.23$ & $0.19$ & $79.90$ & $52.70$ & $51.98$ \\
XMM-Cha\,I-37  & \myrule Glass\,I     & 784  & 1.37 (47) & $=0.94$ & $0.19$ & $3.61$  & $54.99$ & $53.10$ \\
XMM-Cha\,I-47  & \myrule Sz\,30       & 1036 & 1.29 (64) & $=0.21$ & $0.32$ & $1.24$  & $53.41$ & $53.30$ \\
XMM-Cha\,I-48  & \myrule KG\,2001-78  & 190  & 1.27 (12) & $=0.50$ & $0.08$ & $0.20$  & $54.62$ & $53.59$ \\
XMM-Cha\,I-54  & \myrule WX\,Cha      & 905  & 1.35 (33) & $=0.37$ & $0.62$ & $2.28$  & $52.75$ & $52.95$ \\
XMM-Cha\,I-55  & \myrule CHXR-47      & 357  & 0.99 (24) & $=0.88$ & $0.17$ & $1.39$  & $54.06$ & $53.11$ \\
\hline
\end{tabular}
\end{center}
\end{table*}

\subsection{X-ray Temperatures and Spectral Hardness}\label{subsect:kt}

\begin{figure}
\begin{center}
\resizebox{9cm}{!}{\includegraphics{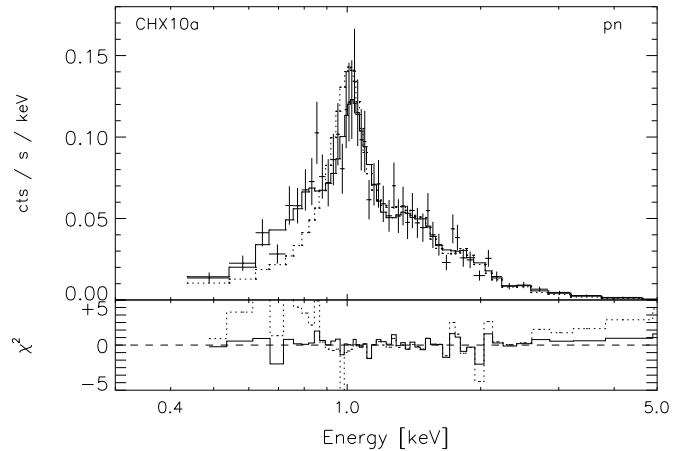}}
\caption{EPIC pn spectrum of CHX\,10a overlaid by the 1-T (dotted histogram) and 2-T (solid histogram)
bestfit model. The lower panel shows the $\chi^2$ residuals. Clearly the 1-T  model does not adequately
describe the data.}
\label{fig:spec_chx10a}
\end{center}
\end{figure}

Among the 25 Cha\,I sources bright enough for the spectral analysis 12  
can be described well by a 1-T model. 
For 13 stars only adding a second temperature component leads to a fit that is acceptable 
on basis of $\chi^2_{\rm red}$ ($< 2$) and upon visual inspection of the residuals. 
As an example we show in Fig.~\ref{fig:spec_chx10a} the pn-spectrum of CHX\,10a, one of
the stars with the highest S/N in this observation, 
together with the best-fit 1-T and 2-T models. Clearly, the 1-T model is not able to provide
a good description of the data in the full energy range. 

In all 2-T models the lower temperature tends to be $\sim 3$\,MK, and the
higher temperature reaches $\sim 10-15$\,MK. The temperatures associated with the 1-T models 
are intermediate between the low and the high energy component of the 2-T models, 
underlining once more 
that an iso-thermal plasma is a simplification of the actual coronal temperature structure. 
This is also obvious by the fact that 
virtually all X-ray bright objects require the 2-T model. 
The temperatures are not very sensitive to changes in $N_{\rm H}$ as we can judge from the
fact that the $kT$ in the 2-T models with free and with fixed column density do not 
differ significantly from each other. However, the $EM$ of the soft component depends 
sensitively on $N_{\rm H}$. 
We note a few cases where $N_{\rm H,X} > N_{\rm H,V}$, and the high absorption in the fit 
with free column density is compensated by 
very high $EM$ of the soft component. Given that these emission measures translate 
into unrealistically high X-ray luminosities, we consider the fits with $N_{\rm H,V}$ 
more reliable. All in all our uniform 2-T models provide a rather stable low-temperature
component with $\sim3$\,MK, but the higher temperature shows a larger spread,
and there is no universal relation between $EM_1$ and $EM_2$, justifying the `definition'
of a typical X-ray spectrum for the Cha\,I stars.  

\begin{figure}
\begin{center}
\resizebox{9cm}{!}{\includegraphics{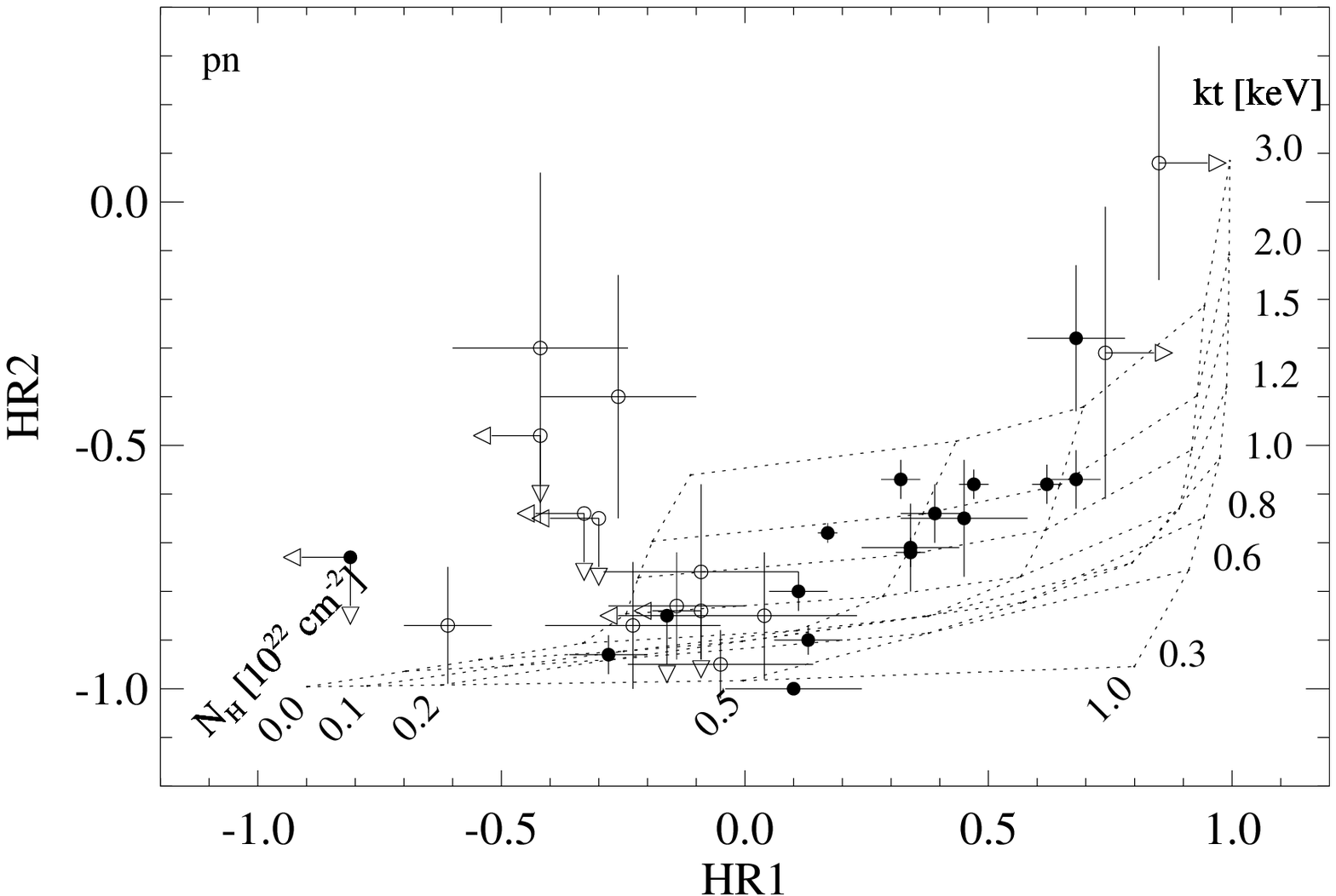}}
\resizebox{9cm}{!}{\includegraphics{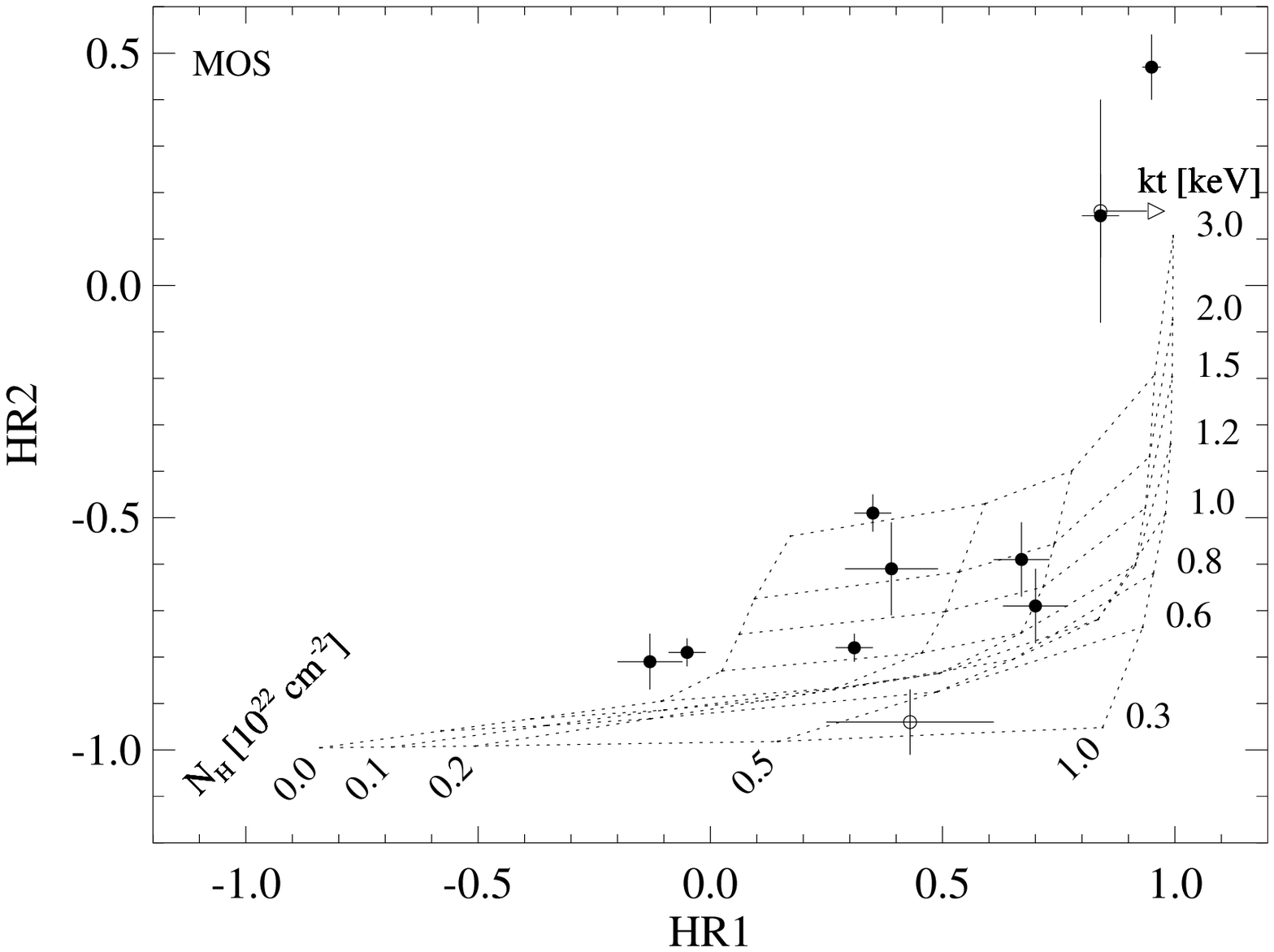}}
\caption{Hardness ratio diagrams for known Cha\,I members in the {\em XMM-Newton} field. 
$HR\,1$ compares the hard\,1 ($1.0-2.4$\,keV) and the soft ($0.3-1.0$\,keV) band, 
and $HR\,2$ compares the hard\,2 ($2.4-7.8$\,keV) and the hard\,1 band. 
Also shown is a grid of column density and temperature corresponding to the hardness ratios derived from the count rates of a 1-T thermal Raymond-Smith model within PIMMS taking account of the different spectral response of pn (top panel) and MOS (bottom panel). Hardness ratios for each source are derived from the number of source counts in the respective energy bands: {\em filled circles} - bright Cha\,I members (spectral sample), {\em open circles} - faint Cha\,I members.} 
\label{fig:hr1_hr2_iden}
\end{center}
\end{figure}

A total of 30 detections are not bright enough for the analysis
of the X-ray spectrum. To obtain an idea about their X-ray temperature we make use
of hardness ratios. The hardness ratio is defined as follows:
$HR = (A-B)/(A+B)$, where $A$ and $B$ are the number of counts in two
adjacent energy bands, and $A$ stands for the higher energy. 
Thus the higher the hardness ratio the harder the spectrum, i.e. the more emission 
is emitted at high energies.
Fig.~\ref{fig:hr1_hr2_iden} shows the hardness ratios computed for 
a grid of $N_{\rm H}$ and $kT$ derived from 1-T Raymond-Smith models 
(\cite{Raymond77.1}) using 
PIMMS\footnote{PIMMS can be accessed via the URL http://asc.harvard.edu/toolkit/pimms.jsp}.
The energy bands are defined as in  Sect.~\ref{subsect:data_red}.
$HR\,1$ compares the hard\,1 and the soft band, and $HR\,2$ the hard\,2 and the
hard\,1 band. 
Since the hardness ratio depends on the spectral response of the detector,
pn and MOS data must be regarded separately. For X-ray sources from the merged data set
we used the pn hardness ratio because of the higher sensitivity with respect to MOS. 
Note that the sensitivity to absorption of the hardness ratio derived from any EPIC detector 
is limited due to the low-energy cutoff at $0.3$\,keV which is urged by the need to avoid 
uncertainties in the calibration of EPIC below this threshold. 

%
%

In Fig.~\ref{fig:hr1_hr2_iden} the observed hardness ratios 
for bright and faint Cha\,I members (i.e. all detections except the unidentified sources) 
are shown. As outlined above most of the bright Cha\,I stars can not be successfully 
fitted by a 1-T model, and their location in the $HR\,1/HR\,2$ diagram is misleading 
due to the simplifying assumption of an iso-thermal plasma 
(see discussion in Sect.~\ref{subsect:lx}). The three X-ray bright objects 
with the largest $HR\,2$ (HD\,97048, HM\,16, and CCE\,98-32) 
are spatially adjacent to each other, 
and among the most absorbed stars in this field. The high temperature present
in their atmospheres is also apparent in their spectra which 
are discussed in Sect.~\ref{subsect:hd97048_group}.   
The faint Cha\,I stars in Fig.~\ref{fig:hr1_hr2_iden} include 
two candidate members, KG\,2001-5 and 2M\,110506+773944.
Although subject to rather large uncertainties
all but four of the X-ray faint objects (UX\,Cha, BYB\,43, CCE\,98-23, and the 2\,MASS source) 
tend to show small $N_{\rm H}$, in agreement with their moderate absorption in the NIR,   
and display coronal temperatures below $\sim 2$\,keV. 
UX\,Cha is a poorly studied variable star for which we derive a visual extinction of
$A_{\rm V} = 3.2$\,mag using 2\,MASS photometry in the $JHK$ color-color diagram. 
The 2\,MASS counterpart
of XMM-Cha\,I-3 may be a PMS star seen through an extinction of $A_{\rm V} \sim 8$\,mag 
(see Fig.~\ref{fig:jh_hk}). 
BYB\,43 was first detected by IRAS (\cite{Baud84.1}). 
Due to its large $K-L$ excess (\cite{Kenyon01.1}) 
and its very strong H$\alpha$ emission (\cite{Comeron04.1}) this object
seems to be a cTTS.  
BYB\,43 has a visual extinction of $A_{\rm V} = 7.6$\,mag 
if transforming the $A_{\rm J}$ given by \citey{Comeron04.1}
with the RL85 extinction law. Similarly, we find 
$A_{\rm V} = 30$\,mag for CCE\,98-23 (\cite{Gomez03.1}). 
Bearing in mind their high extinction, the (apparently) 
hard X-ray spectrum of these stars does not come as a surprise.

\subsection{X-ray Luminosities}\label{subsect:lx}

We derive individual absorption corrected X-ray luminosities for each Cha\,I member
and the X-ray detected candidate members. 
We chose to compute $L_{\rm x}$ in the $0.4-2.5$\,keV band to allow for 
comparison with earlier {\em ROSAT} measurements. 
The spectra revealed that the Cha\,I stars emit most of their X-rays below
$\sim 2$\,keV. 

In view of the inhomogeneous statistics of the X-ray spectra 
it is impossible to determine $L_{\rm x}$ for all stars in the same way. 
Below we describe the methods used to derive $L_{\rm x}$. 
The luminosities for the $0.4-2.5$\,keV band are listed in Table~\ref{tab:lx},  
where the method is indicated by the flag in column~3. 
The distance assumed is $160$\,pc (\cite{Whittet97.1}). 
The photon extraction area for the spectral analysis contains between 
$\sim 70 - 85$\,\% of the source photons, and accordingly a PSF correction factor
was applied to make up for the missing counts.
%
%
\begin{table}
\begin{center}
\caption{X-ray luminosities in the $0.4-2.5$\,keV band for $d=160$\,pc, X-ray to bolometric luminosity ratio,
and probability for variable source according to the KS-test. 
The flag in column~3 indicates the method used to derive $L_{\rm x}$: `S' - Spectral fit with
2-T model and column density fixed to the value expected for the galactic 
$N_{\rm H}-A_{\rm V}$ relation, 
'L' - Absorption from the literature and $\log{T}\,[K] = 6.95$, 'N' - Absorption
newly derived in this paper from $JHK$ colors and $\log{T}\,[K] = 6.95$.}
\label{tab:lx}
\begin{tabular}{lrrrr} \hline
Designation & $\log{L_{\rm x}}$ & Flag        & $\log{(L_{\rm x}/L_{\rm bol})}$ & KS prob. \\
            & [erg/s]          & $L_{\rm x}$ &  & \\
\hline
\multicolumn{5}{c}{Spectroscopically confirmed Cha\,I members} \\
\hline
KG\,2001-19    &  $28.67$ & L     &  $-3.66$ & $-$     \\
CHXR-21        &  $30.83$ & S     &  $-2.40$ & 95\,\%  \\
HM\,19         &  $29.44$ & L     &  $-3.52$ & $-$     \\
VW\,Cha        &  $30.97$ & S     &  $-3.13$ & 100\,\% \\
CHXR-74        &  $30.30$ & S     &  $-2.57$ & $-$     \\
BYB\,18        &  $29.36$ & S     &  $-3.25$ & 100\,\% \\
KG\,2001-96    &  $28.90$ & L     &  $-3.45$ & $-$     \\
HD\,97048      &  $29.58$ & S     &  $-5.72$ & $-$     \\
CCE\,98-19     &  $30.92$ & S     &  $-2.41$ & $-$     \\
CHXR-73        &  $30.49$ & S     &  $-2.31$ & 100\,\% \\
CHXR-76        &  $29.92$ & S     &  $-2.62$ & $-$     \\
CHXR-26        &  $31.27$ & S     &  $-2.35$ & 100\,\% \\
CHXR-47        &  $30.77$ & S     &  $-3.75$ & $-$     \\
CHX-10a        &  $30.56$ & S     &  $-3.64$ & 99\,\%  \\
CHXR-15        &  $29.56$ & S     &  $-3.10$ & $-$ \\
BYB\,43        &  $29.08$ & L     &  $-3.88$ & 100\,\% \\
Ced\,110\,IRS-2&  $31.16$ & S     &  $-3.69$ & 100\,\% \\
WX\,Cha        &  $30.17$ & S     &  $-3.35$ & $-$ \\
CHXR-78\,C     &  $28.96$ & L     &  $-3.44$ & $-$     \\
CCE\,98-23     &  $30.33$ & L     &  $-3.32$ & $-$     \\
Sz\,30         &  $30.62$ & S     &  $-2.86$ & 99\,\%  \\
CHXR-25        &  $29.82$ & S     &  $-2.63$ & 97\,\%  \\
CHXR-22\,E     &  $30.94$ & S     &  $-2.87$ & $-$     \\
HM\,15         &  $30.47$ & S     &  $-3.18$ & $-$     \\
CCE\,98-32     &  $31.06$ & S     &  $-2.92$ & 100\,\% \\
HM\,16         &  $31.00$ & S     &  $-3.31$ & $-$     \\
LH$\alpha$332-17& $31.09$ & S     &  $-3.70$ & 97\,\%  \\
Glass\,I       &  $31.79$ & S     &  $-2.65$ & $-$     \\
CHXR-20        &  $30.03$ & S     &  $-3.49$ & $-$     \\
ChaH$\alpha$1  &  $28.25$ & L     &  $-3.39$ & $-$     \\
ChaH$\alpha$2  &  $28.49$ & L     &  $-3.52$ & $-$     \\
ChaH$\alpha$3  &  $29.11$ & S     &  $-3.03$ & $-$     \\
ChaH$\alpha$4  &  $28.41$ & L     &  $-3.85$ & $-$     \\
ChaH$\alpha$5  &  $29.21$ & L     &  $-2.98$ & $-$     \\
ChaH$\alpha$6  &  $28.91$ & L     &  $-3.12$ & $-$     \\
ChaH$\alpha$7  &  $28.27$ & L     &  $-3.14$ & $-$     \\
ChaH$\alpha$8  &  $28.44$ & L     &  $-3.39$ & 99\,\%  \\
ChaH$\alpha$9  & $<27.92$ & L     & $<-3.33$ & $-$ \\
ChaH$\alpha$10 & $<27.78$ & L     & $<-3.54$ & $-$ \\
ChaH$\alpha$11 & $<27.67$ & L     & $<-3.48$ & $-$ \\
ChaH$\alpha$12 & $<28.19$ & L     & $<-3.81$ & $-$ \\
ChaH$\alpha$13 &  $29.02$ & L     &  $-3.51$ & 97\,\%  \\
IR\,Nebula     & $<29.16$ & L     & $<-4.77$ & $-$ \\
\hline
\multicolumn{5}{c}{Photometric candidates} \\
\hline
UX\,Cha        &  $29.53$ & N     &  $-3.74$ & $-$      \\
KG\,2001-78    &  $30.56$ & S     &  $-4.56$ & $-$      \\
KG\,2001-5     &  $28.86$ & L     &  $-3.93$ & $-$      \\
2M\,110506-773944&$29.51$ & N     &  $-3.34$ & $-$    \\
\hline
\end{tabular}
\end{center}
\end{table}

\subsubsection{Bright (Spectral) Sample}\label{subsubsect:lx_bright}

For sources from the bright (spectral) sample $L_{\rm x}$ results 
directly from the X-ray spectrum. As discussed above some doubts remain about
the best fit due to the existence of several local minima in the $\chi^2$ space. 
But the fits with column density based on the opt./IR extinction are preferred over 
the ones with free column density. 
Table~\ref{tab:lx} shows the X-ray luminosities $L_{\rm x,V}$ computed from the  
2-T best fits with fixed $N_{\rm H,V}$.  

Here we want to add some notes to recall the difficulties in establishing reliable X-ray 
luminosities: 
When confronting the resulting luminosities $L_{\rm x,V}$ with those where the column density 
was a free fit parameter ($L_{\rm x,X}$) 
the following pattern emerges: 
\begin{itemize}
\item For the sources where the fit with free $N_{\rm H,X}$ 
requires two temperatures the luminosities $L_{\rm x,X}$ 
are similar to those from the 2-T fit with fixed $N_{\rm H,V}$, except where $N_{\rm H,X}$ 
is significantly higher than $N_{\rm H,V}$. In these latter cases 
$L_{\rm x,X}$ is higher than $L_{\rm x,V}$.  
\item For sources that can be described with a 1-T model and free $N_{\rm H,X}$ the 
luminosities are systematically lower than those of the 2-T fit with fixed $N_{\rm H,V}$. 
We attribute this to the mis-estimate of the temperature and emission measure 
distribution that goes along with the 1-T model (see discussion in Sect.~\ref{subsect:kt}). 
\end{itemize}

\subsubsection{Faint Sample}\label{subsubsect:lx_faint}

Since the analysis of the bright sample did not result in an obvious, typical 
spectral shape for a 2-T model, 
for stars from the faint sample we must rely on assumptions made for 
the parameters of a 1-T model, and determine the X-ray luminosities 
from the count rate with help of PIMMS. 
%
%
\begin{figure}
\begin{center}
\resizebox{9cm}{!}{\includegraphics{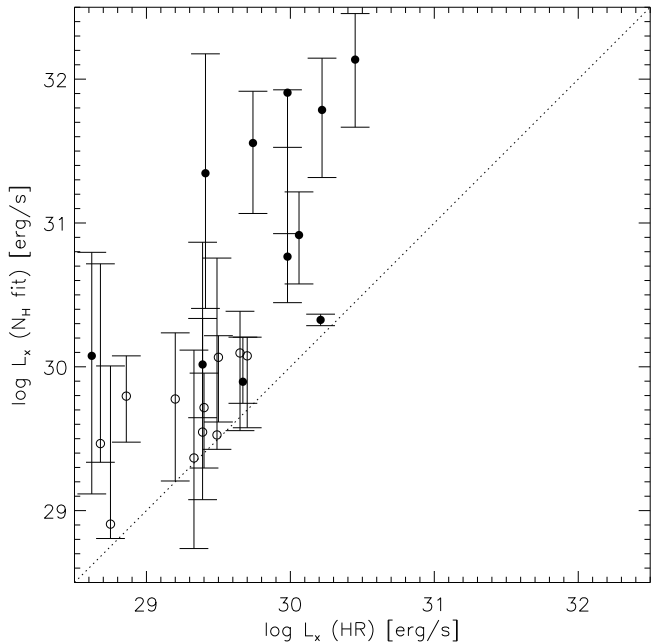}}
\caption{Comparison of the X-ray luminosities of X-ray bright Cha\,I objects derived with two
different approaches: {\em x-axis} - using the $HR$ to guess the spectral model which is 
converted to X-ray flux using PIMMS, and {\em y-axis} - from the X-ray spectrum using XSPEC. 
Filled plotting symbols denote stars which can not be described by a 1-T model.  
As can be seen, for these stars the $HR$ tend to provide a wrong estimate of $L_{\rm x}$.}
\label{fig:lx_hr_xspec}
\end{center}
\end{figure}
We resort to the hardness ratio to guess the spectral model on an individual basis. 
To check the validity of this approach, we make use of the bright sample,
for which we compare the luminosities derived with PIMMS from the 1-T model based on the 
hardness ratios to the luminosities $L_{\rm x,X}$ extracted from the spectrum. 
Note that the comparison is based on the spectral fits with free $N_{\rm H,X}$.  
The reason to use the free-$N_{\rm H,X}$ fits here
is because we want to demonstrate the effects of different model assumptions. 
(Recall, however, that the 2-T fits with fixed $N_{\rm H,V}$ were performed in the most 
possible uniform way, and are considered more realistic as explained above.) 

The X-ray luminosities derived from the spectral fits on the one hand, 
and from the hardness ratios on the other hand are shown in Fig.~\ref{fig:lx_hr_xspec}. 
The uncertainties in the X-ray luminosities derive from 
the uncertainties in $N_{\rm H,X}$. 
Good agreement between the XSPEC fit results and the PIMMS estimates is found for 
the luminosities of stars the spectra of which can be described by an iso-thermal model. 
The discrepancies in Fig.~\ref{fig:lx_hr_xspec} for the stars that require a 2-T model is 
obvious, underlining that the use of the 1-T grid in the $HR-$diagram is inappropriate. 
Since virtually all the bright sources, i.e. the ones with good
statistics, require two temperatures, we caution that {\em any} X-ray luminosity derived
from an iso-thermal model may be wrong. The non-linearity of the count-to-energy conversion
factor (evident by the curved shape of the model grid in Fig.~\ref{fig:hr1_hr2_iden})
leads to a systematic underestimation of $N_{\rm H}$ in a 1-T model, and consequently
the unabsorbed flux is also underestimated. 
Despite these drawbacks we must nevertheless use this approach for the faint X-ray sources, 
but be careful with the interpretation of the absolute values for the luminosity. 

The location of the objects in the upper left of the range of $N_{\rm H}$ and 
$kT$ spanned by the grid in Fig.~\ref{fig:hr1_hr2_iden} is probably the result
of uncertain background level. All these faint sources are M-type stars and
brown dwarfs. For these objects we 
assume a coronal temperature of $0.8$\,keV (representative 
for the X-ray spectrum of ChaH$\alpha$\,3, 
the latest-type object from the bright sample),  
and a column density corresponding to their individual visual extinction. 
The luminosity of ChaH$\alpha$\,8, which is not detected in any of the individual
detectors but only in the merged pn + MOS image, is computed in the same way, and 
making use of the response of MOS within PIMMS. 

We computed upper limits for the count rates of non-detected Cha\,I members  
using the tabulated upper limit counts derived by \citey{Gehrels86.1}. 
The upper limit to the flux and luminosity is derived for $kT = 0.8$\,keV using the
visual/IR absorption if known, else the value estimated from the $J-H / H-K$ diagram
(see Sect.~\ref{subsect:extinc}).

\section{Time Series Analysis}\label{sect:lcs}

For the temporal analysis we used the same photon extraction area as for the spectral 
analysis. 
To check for variability we applied the KS-test to the photon arrival times of all 
X-ray sources in the field after removing data gaps. Variability was examined only within
the first 14.5\,ksec of the observation, because the end of the data set is contaminated by
strong background flaring resulting in frequent and long periods without data, interrupted
by only short intervals of `good-times'. 
The KS-test was performed in the broad, the soft and the hard\,1 band. 
In the broad band a total of $16$ (out of $58$) X-ray sources are variable with 
probability $>95$\,\%.
Most variable sources ($14$) are Cha\,I members or candidates. 
Most of the variability in the broad band 
can be attributed to the hard\,1 band, while the emission in the soft band is more steady. 
The probabilities for sources which are variable at a significance level $>95$\,\% 
are summarized in the last column of Table~\ref{tab:lx}.

\begin{figure}
\begin{center}
\parbox{9cm}{\resizebox{9cm}{!}{\includegraphics{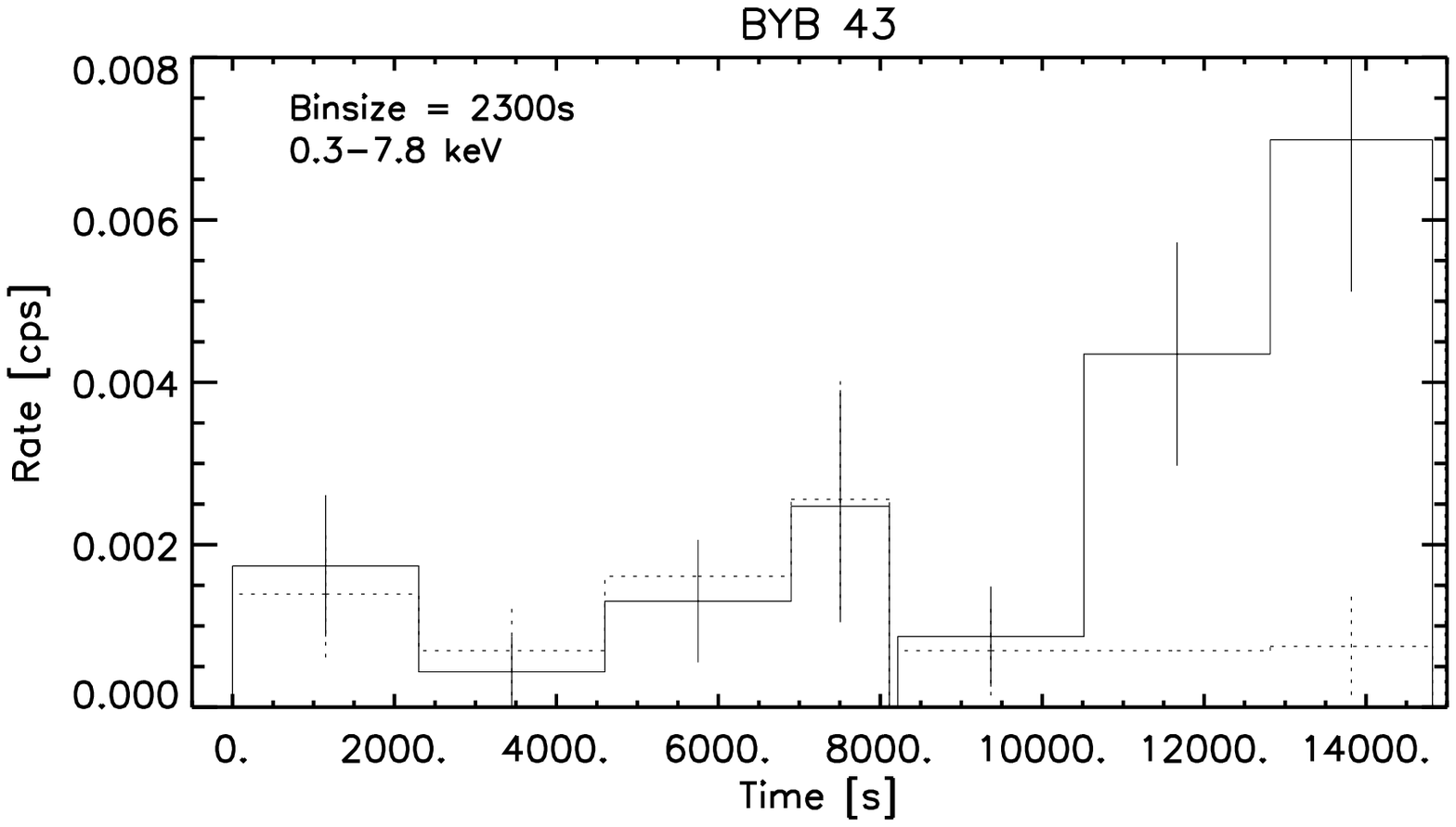}}}
\parbox{9cm}{\resizebox{9cm}{!}{\includegraphics{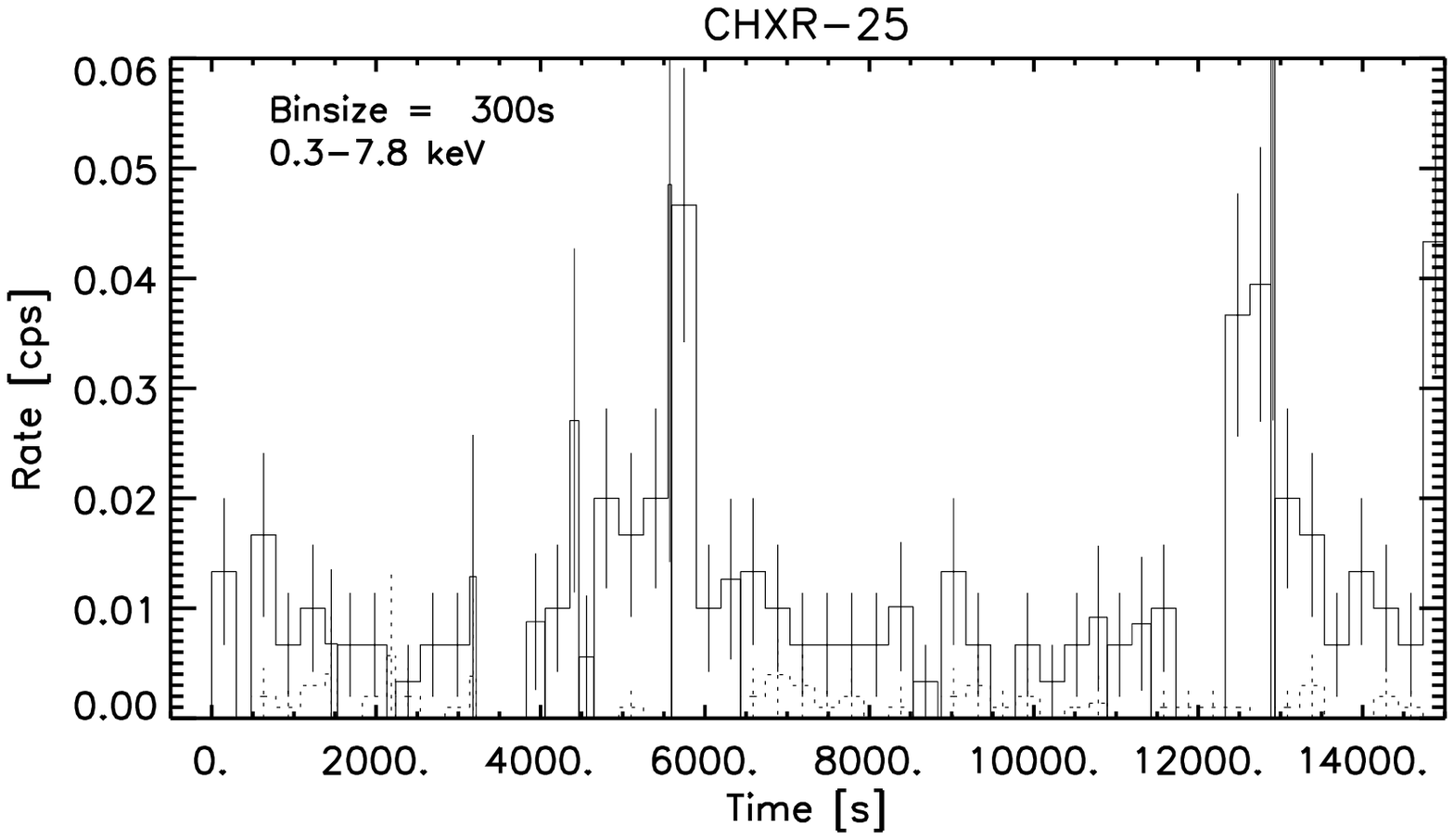}}}
\caption{Broad band EPIC lightcurve of the cTTS BYB\,43 and the wTTS CHXR-25 displaying flares. The dotted line represents the background extracted from an adjacent source free region. The background has been scaled to the size of the extraction area for the source photons.} 
\label{fig:lc_varstars}
\end{center}
\end{figure}

Broad, soft, and hard\,1 band lightcurves were binned for all variable sources and all
other sources with at least $>~150$~counts. 
Some of the variable stars seem to have undergone irregular variability, 
others have shown flares. However, there is no evidence for such dramatic outbursts, 
as often reported for low-mass PMS stars 
(e.g. \cite{Skinner97.1}, \cite{Tsuboi98.1}, \cite{Stelzer00.2}, \cite{Hamaguchi00.1}). 
We display two of the most interesting lightcurves in Fig.~\ref{fig:lc_varstars}. 
These examples show that flares occur on cTTS as well as on wTTS. The cTTS BYB\,43 is 
detected for the first time in X-rays, and only because of the flare at the end of the 
observation. The wTTS CHXR-25 suffered no less than three outbursts during the observing 
time, amounting to an unusually high flare rate. 
A number of other interesting variable sources will be discussed in the next section.

\section{Individual (Groups of) X-ray Sources}\label{sect:indiv}

\subsection{The HD\,97048 Cluster}\label{subsect:hd97048_group}

HD 97048 is a Herbig Ae/Be star with spectral type 
B9-A0ep+sh (\cite{vandenAncker98.1}).  
It is extended in the mid-IR (Prusti et al. 1994) indicating the
presence of a dusty envelope. 
HD\,97048 is one of two stars in Cha\,I with a {\em HIPPARCOS} parallax.
This measurement puts the star at $175^{+27}_{-21}$\,pc (\cite{Wichmann98.1}). 

F93 have identified their X-ray source CHXR-29 with
HD\,97048. But \citey{Zinnecker94.1} argue that the X-ray source is the nearby
TTS HM\,16 (=Sz\,22). In the much longer {\em ROSAT} pointing
presented by \citey{Zinnecker94.1} HD\,97048 is detected as a weak X-ray 
source, marginally resolved from HM\,16. 
Now {\em XMM-Newton} resolved these two stars for the first time clearly.
In addition, a new X-ray source is discovered close to HM\,16. This latter source
corresponds to CCE\,98-32, an M2-type star. CCE\,98-32 is identical with a visual companion 
to HM\,16 mentioned by \citey{Ghez97.1}. The separation ($16.4^{\prime\prime}$) 
and position angle ($80.2^\circ$) we measure between the X-ray sources are in agreement 
with the values provided by \citey{Ghez97.1}.

HD\,97048 is one of the hardest sources in Fig.~\ref{fig:hr1_hr2_iden}.
A 1-T model does not provide a satisfactory description of the X-ray spectrum 
($\chi^2_{\rm red} > 2$). The parameters of the 2-T model 
indicate strong absorption, a dominating soft component, 
and a weaker, poorly constrained high-energy tail. 

The origin of X-ray emission from HAeBe stars remains mysterious. 
Several emission meachanism including coronal activity like on late-type
stars, strong stellar winds, and unresolved T Tauri companions are being discussed. 
HD\,97048 is not known to be a binary: A Speckle imaging search (\cite{Ghez97.1}), 
a spectroscopic search for binarity (\cite{Corporon99.1}),
and spectro-astrometry (\cite{Bailey98.1}, \cite{Takami03.1}) 
have all provided negative results. 
Therefore, the observed X-ray emission is most likely intrinsic to HD\,97048. 
Its origin in a wind is improbable on basis of its X-ray properties:
X-ray emission from stellar wind sources is much softer 
($kT \leq 0.5$\,keV; \cite{Berghoefer96.1}) than observed here. 
HD\,97048 was not detected in a VLA survey,
unlike several other HAeBe stars for which \citey{Skinner93.1} have suggested 
a thermal wind as the origin of their radio emission. 

The two X-ray sources corresponding to HM\,16 and CCE\,98-32
are too closely spaced for our standard spectral photon extraction procedure.
In order to correct for possible contributions by the respective neighboring   
source we extracted the background spectrum from an annulus centered on the position of the
adjacent contaminating source but exclude the area of the source under consideration. 
The column density inferred from $A_{\rm J}$ is $N_{\rm H}>10^{22}\,{\rm cm^{-2}}$ for both stars,
consistent with the almost complete lack of photons in the soft band;   
see also the location of CCE\,98-32 and HM\,16 at the upper right in the 
hardness ratio diagram (Fig.~\ref{fig:hr1_hr2_iden}). 
As a result the low-temperature component is poorly constrained by the spectral fit,  
and the numbers for the emission measures and X-ray luminosities of HM\,16 and CCE\,98-32 
in Table~\ref{tab:specparams_nhlit} and~\ref{tab:lx} are uncertain by up to one order
of magnitude. 

\begin{figure}
\begin{center}
\resizebox{9cm}{!}{\includegraphics{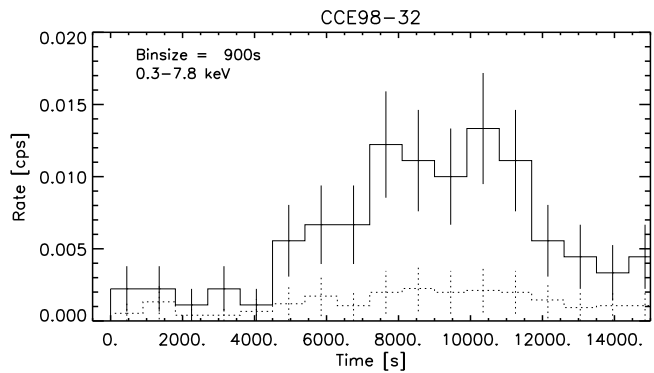}}
\caption{Broad band MOS\,2 lightcurve of CCE\,98-32. The lower curve is the background lightcurve, with count rate scaled corresponding to the extraction area.}
\label{fig:lc_denis32}
\end{center}
\end{figure}

According to the KS-test CCE\,98-32 is variable in the broad band, but HM\,16 and 
HD\,97048 are not. 
The negative result of the KS-test in the soft and hard\,1 band is not surprising, 
because these stars emit -- in contrast to the rest of the Cha\,I sample 
-- most of their X-rays at energies $>1$\,keV. 
Inspection of the broad band lightcurve of CCE\,98-32 (Fig.~\ref{fig:lc_denis32}) 
shows a smooth and slow increase and subsequent decline of the count rate. 

\subsection{KG\,2001-78}\label{subsect:spirex78}

KG\,2001-78 is a poorly studied NIR candidate member of Cha\,I (see \cite{Kenyon01.1}).
From its position in the $JHK$
color-color diagram we find that this is an A-type star. 
But, in contrast to HD\,97048, no signs for a NIR excess are observed. 
Among the X-ray sources discussed  here it stands out as particularly soft (the hard
component is clearly missing) and little absorbed; see Table~\ref{tab:specparams_nhlit}. 
In Fig.~\ref{fig:hr1_hr2_iden}
KG\,2001-78 is the only object in the bright sample that appears to the left of the unabsorbed
1-T model. We speculate that this might be an indication that its X-ray emission is of
different origin. Indeed, its soft spectrum and lack of variability 
are typical characteristics of wind-driven X-ray sources. 
The marked difference between the X-ray properties of the only two intermediate mass
stars in the FOV (KG\,2001-78 and HD\,97048) is puzzling, because 
according to their position in the HRD they have similar mass and age. 
However, as judging from the lack of a NIR excess KG\,2001-78 may be more evolved,
and activity seems to have ceased, 
leaving the wind as only driving mechanism for the X-ray emission.

\subsection{Binary T Tauri Stars}\label{subsect:binary}

The {\em XMM-Newton} FOV comprises some binary systems. 
LH$\alpha$332-17, WX\,Cha, and Glass\,I are visual binaries according to 
\citey{Ghez97.1} and \citey{Chelli88.1}. 
VW\,Cha is a triple system (\cite{Brandeker01.1}) with a possible forth
companion (\cite{Ghez97.1}). 
The separation in these multiples ranges from 
the sub-arcsecond regime up to $\sim 5^{\prime\prime}$, such that all of them are
unresolved with {\em XMM-Newton}. 

The TTS Sz\,23 is separated from VW\,Cha by only $16^{\prime\prime}$, 
and confused with the bright X-ray source at the position of VW\,Cha in the
{\em XMM-Newton} image. 

LFH96 presented two counterparts to the X-ray source CHXR-22. CHXR-22\,E
is an early M-star, and CHXR-22\,W a heavily obscured star with slightly
earlier spectral type. Their separation is only $9.5^{\prime\prime}$, such 
that they are represented by a single X-ray source with {\em XMM-Newton}.  
However, the X-ray position is much closer to the Eastern component 
(see Table~\ref{tab:x-sources}). 

Two of the VLM objects in this field have been listed as possible binaries
by \citey{Neuhaeuser02.1}. While the companion candidate ChaH$\alpha$\,5 was 
later revealed to be a background object (\cite{Neuhaeuser03.1}), 
Cha\,H$\alpha$\,2 might be a binary with a separation of 
$0.2^{\prime\prime}$.

\subsection{Cha IRN}\label{subsect:irn}

The bipolar reflection nebula Chamaeleon IR nebula is the only 
spectroscopically confirmed
Cha\,I member in the {\em XMM-Newton} FOV which is not detected in X-rays. 
The central object is hidden behind $A_{\rm V} \sim 10$\,mag, and was believed to 
be an A7....K7 star with an edge-on disk based on IR photometry (\cite{Cohen84.1}). 
But a recent NIR spectrum
provided a spectral type of M5 (\cite{Gomez03.1}). The disk may
prevent us from seeing the X-ray emission generated by the young star in the
center.

\subsection{VLM H$\alpha$ Stars and Brown Dwarfs}\label{subsect:bds}

\begin{figure}
\begin{center}
\resizebox{9cm}{!}{\includegraphics{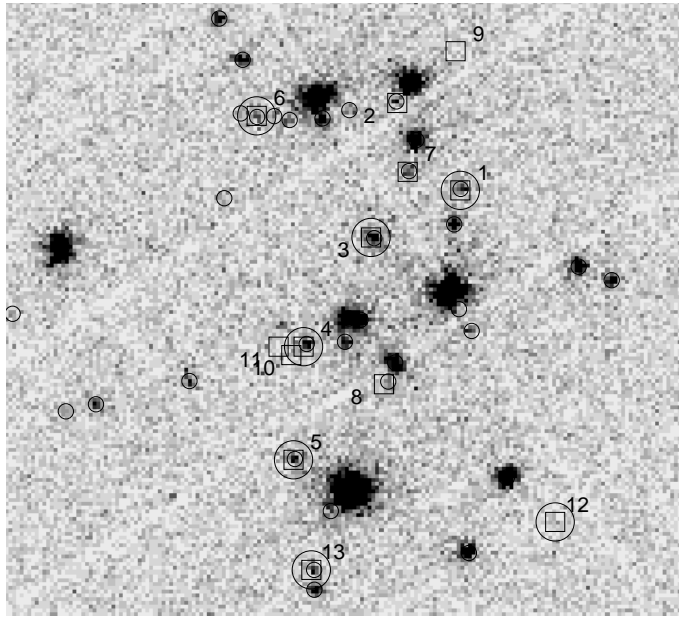}}
\caption{Merged image of EPIC pn, MOS\,1 and MOS\,2 showing all VLM H$\alpha$ emitters in Cha\,I from the survey by CNK00. The IR positions of Cha\,H$\alpha$\,1...13 are indicated by squares and numbers. Small circles are {\em XMM-Newton} sources, and large circles represent the {\em ROSAT} PSPC error boxes of $25^{\prime\prime}$ radius at the IR position of those ChaH$\alpha$ objects detected with {\em ROSAT}; no {\em ROSAT} X-ray positions are given in the literature.}
\label{fig:bdimage}
\end{center}
\end{figure}

CNK00 list 7 of the 13 VLM H$\alpha$ emitters in Cha\,I as X-ray detections. 
As mentioned in Sect.~\ref{sect:nature} we detected 9 of these objects with
{\em XMM-Newton}. With the exception of ChaH$\alpha$\,12,
which is not detected with {\em XMM-Newton}, all {\em ROSAT} sources are confirmed. 
{\em XMM-Newton} revealed for the first time X-ray emission from ChaH$\alpha$\,2, 
ChaH$\alpha$\,7 and ChaH$\alpha$\,8.
These three objects may have escaped detection with {\em ROSAT} because they could not be 
resolved from nearby bright TTS. 
ChaH$\alpha$\,4 appeared as an elongated object in the
{\em ROSAT} image. With the improved spatial resolution of {\em XMM-Newton} 
we can separate it 
from ChaH$\alpha$\,10 and ChaH$\alpha$\,11. We find that these latter two are not detected,
and the only X-ray source can clearly be attributed to ChaH$\alpha$\,4.  
A visual comparison of the {\em ROSAT} and {\em XMM-Newton} detections is given in 
Fig.~\ref{fig:bdimage}.

Only one of the objects at the substellar borderline, ChaH$\alpha$\,3, 
which was the prime target of this pointing,
has enough statistics to attempt modeling the X-ray spectrum. 
Analogous to (many of) the higher-mass Cha\,I members with similar number of counts 
ChaH$\alpha$\,3 can be described with 
a single-temperature MEKAL model and a photo-absorption term. 
The X-ray temperature ($\sim 9$\,MK) is indistinguishable from those of the 
higher-mass TTS. But a 2-T model with column density fixed on the values calculated from
the opt./IR extinction fits the data equally well, and provides typical parameters
for late-type stars (see Table~\ref{tab:specparams_nhlit} and Table~\ref{tab:lx}). 
For the remaining detected ChaH$\alpha$ objects the hardness ratios indicate comparatively
soft X-ray spectra, i.e. moderate X-ray temperature and low absorption. 
But this may be a selection effect: All VLM objects in Cha\,I discovered so far 
are located in regions of low extinction. 

The VLM stars and brown dwarfs in Cha\,I do also not appear to differ from the higher 
mass stars in terms of their variability: 
According to the KS-Test 2 out of 9 detected objects with spectral type 
later than M4 are variable, while 12 out of 29 higher-mass stars are variable. 
Therefore, the variability of the VLM objects seems to be similar to that of 
the higher-mass TTS. 
This is in contrast to the observation of old brown dwarfs in the field: The only
field brown dwarf detected so far in X-rays was observed during a flare, but later
observations found only a very restrictive upper limit to its X-ray luminosity 
(\cite{Rutledge00.1}, \cite{Martin02.1}).

\begin{figure}
\begin{center}
\parbox{9cm}{\resizebox{9cm}{!}{\includegraphics{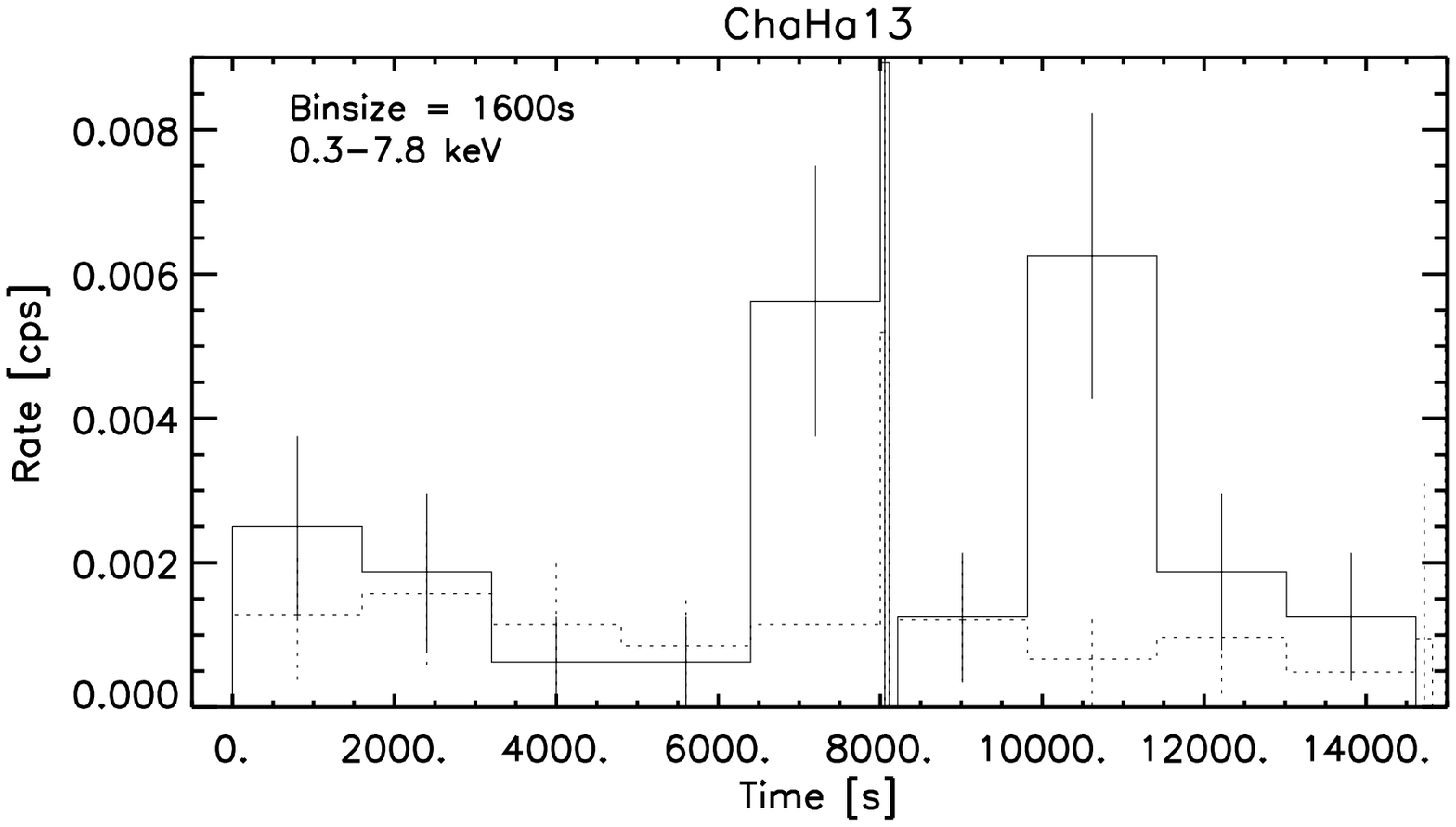}}}
\parbox{9cm}{\resizebox{9cm}{!}{\includegraphics{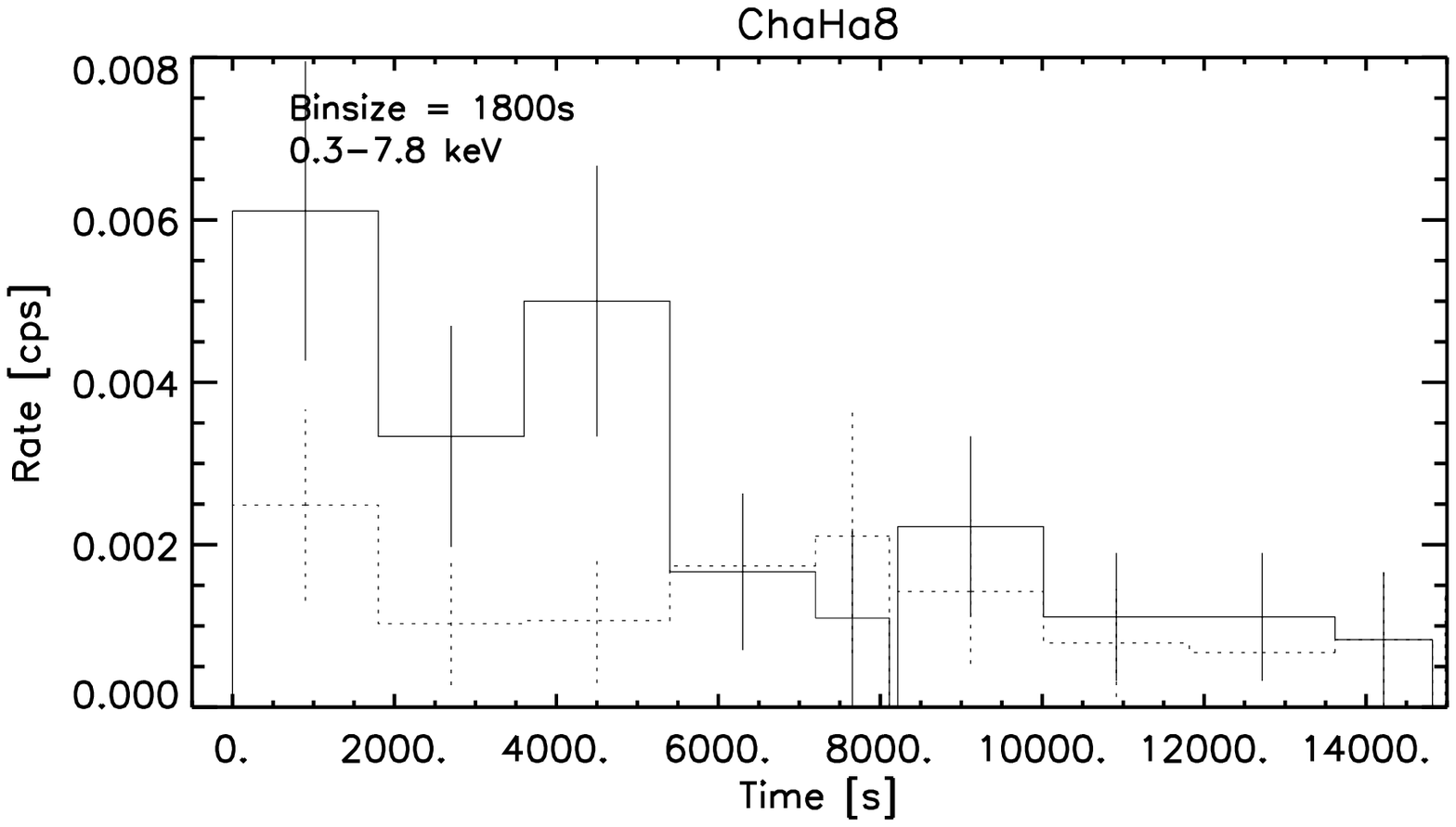}}}
\parbox{9cm}{\resizebox{9cm}{!}{\includegraphics{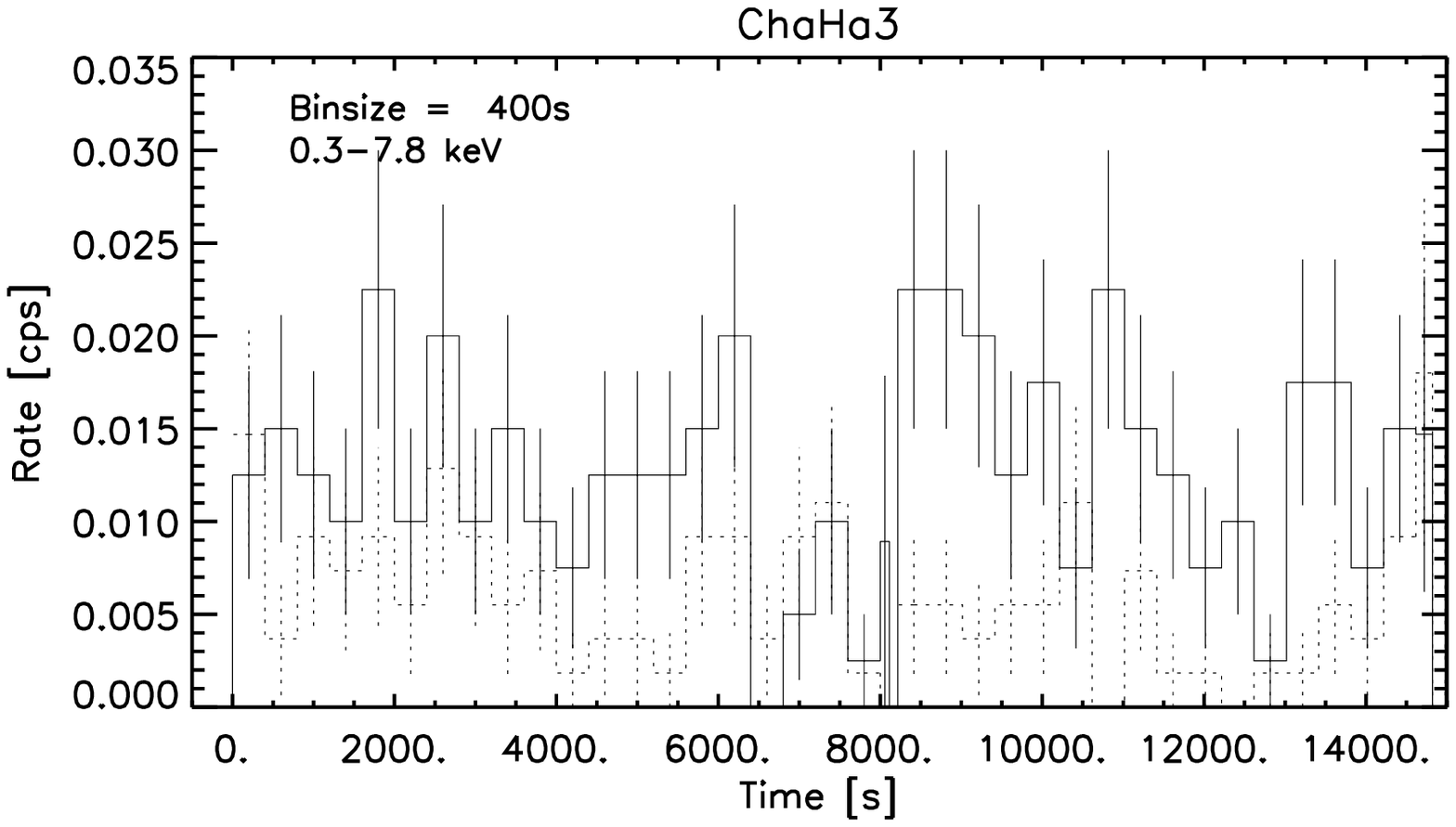}}}
\caption{Broad band X-ray lightcurves of three VLM objects in Cha\,I: ChaH$\alpha$\,13 and ChaH$\alpha$\,8 barely emerge above the background outside flares, while ChaH$\alpha$\,3 is clearly visible throughout the whole observing time.}
\label{fig:lc_vlm}
\end{center}
\end{figure}
However, we point out that the two strongly variable VLM stars ChaH$\alpha$\,13 
(spectral type M5) and ChaH$\alpha$\,8 (spectral type M6.5) have undergone flares 
(Fig.~\ref{fig:lc_vlm}). For ChaH$\alpha$\,13 the lightcurve shows that most of the 
source photons are concentrated in two distinct events near the middle of the 
observation. The duration of these events
$\sim 1/2$\,h is typical for X-ray flares on late-type stars (\cite{Stelzer00.1}).
ChaH$\alpha$\,8 has flared at the beginning of the observation. 
Outside these events both sources emerge barely above the background level. 
In contrast the X-ray brightest of the VLM objects, ChaH$\alpha$\,3,  
is clearly detected throughout the full observing time.

\subsection{Unidentified X-ray sources}\label{subsect:uiden}

\begin{table}
\begin{center}
\caption{Best fit parameters of an absorbed power-law model for X-ray bright 
unidentified X-ray sources in the Cha\,I {\em XMM-Newton} field.}
\label{tab:specparams_uiden}
\begin{tabular}{llrrr} \hline
Designation & \myrule $\chi^2_{\rm red}$ & \multicolumn{1}{c}{$N_{\rm H}$\,[${\rm 10^{22}}$} & \multicolumn{1}{c}{$\Gamma$} & \multicolumn{1}{c}{$f_{\rm x}$} \\    
            & \myrule (d.o.f.)           & ${\rm cm^{-2}}$]                                  &                              & [${\rm erg/cm^2/s}$]            \\
\hline        
XMM-Cha\,I-18 & \myrule 0.91 (3)  & 3.4  & 2.76 & $1.6 \times 10^{-13}$  \\
XMM-Cha\,I-35 & \myrule 0.89 (25) & 0.91 & 1.47 & $3.4 \times 10^{-13}$  \\
XMM-Cha\,I-44 & \myrule 0.80 (6)  & 0.73 & 1.73 & $4.0 \times 10^{-14}$  \\
\hline
\end{tabular}
\end{center}
\end{table}

For four X-ray sources not associated with any Cha\,I star 
or candidate from the literature we found a 2\,MASS counterpart that
seems not to be a cloud member according to its position in the HRD (see
Fig.~\ref{fig:hrd_tracks} and Table~\ref{tab:opt_lx}). 
These objects lie slightly below the zero-age MS, and are underluminous
by $\sim\,1$ order of magnitude with respect to Cha\,I members of similar effective
temperature. It may be worth mentioning that \citey{Comeron03.1} identified young stars 
in the Lupus star forming region 
with bolometric luminosity lower than expected for their age or, in other words, 
stars that according to their position in the HR diagram appear older than expected.
\citey{Hartmann97.1} showed that strong accretion may accelerate stellar evolution, and,
indeed, the underluminous objects in Lupus display unusually strong H$\alpha$ emission. 
If we take X-ray emission as a proxy for youth, the same mechanism may be at work in the
2\,MASS objects in Cha\,I. However, we lack information on their accretion properties. 
Moreover, the NIR photometry used to place them in the HR diagram introduced large 
uncertainties in both spectral type, i.e. effective temperature, and luminosity (via extinction). 
Conclusions about their nature that go beyond our speculative remarks 
require a spectroscopic investigation. 

Only one X-ray source could be identified with a known extragalactic object, 
the BL\,Lac source WGA\,J1108.1-774 (\cite{Perlman98.1}).
Another 11 X-ray sources remain without optical/IR counterpart. 

Within this sample of X-ray sources not confirmed to be Cha\,I members
three are bright enough for the analysis of the spectrum: the BL\,Lac object,
and two of the unidentified X-ray sources. 
For all of them a thermal model results in unusually high and poorly constrained
X-ray temperature, although the statistics of the spectrum are comparable to that for some 
of the known Cha\,I stars in the field. 
We fitted these spectra alternatively with an absorbed power law. The power law indices are 
between $1.5 ...2.7$ (see Table~\ref{tab:specparams_uiden}), which is typical for 
extragalactic sources (\cite{Tozzi01.1}). 

\begin{figure}
\begin{center}
\resizebox{9cm}{!}{\includegraphics{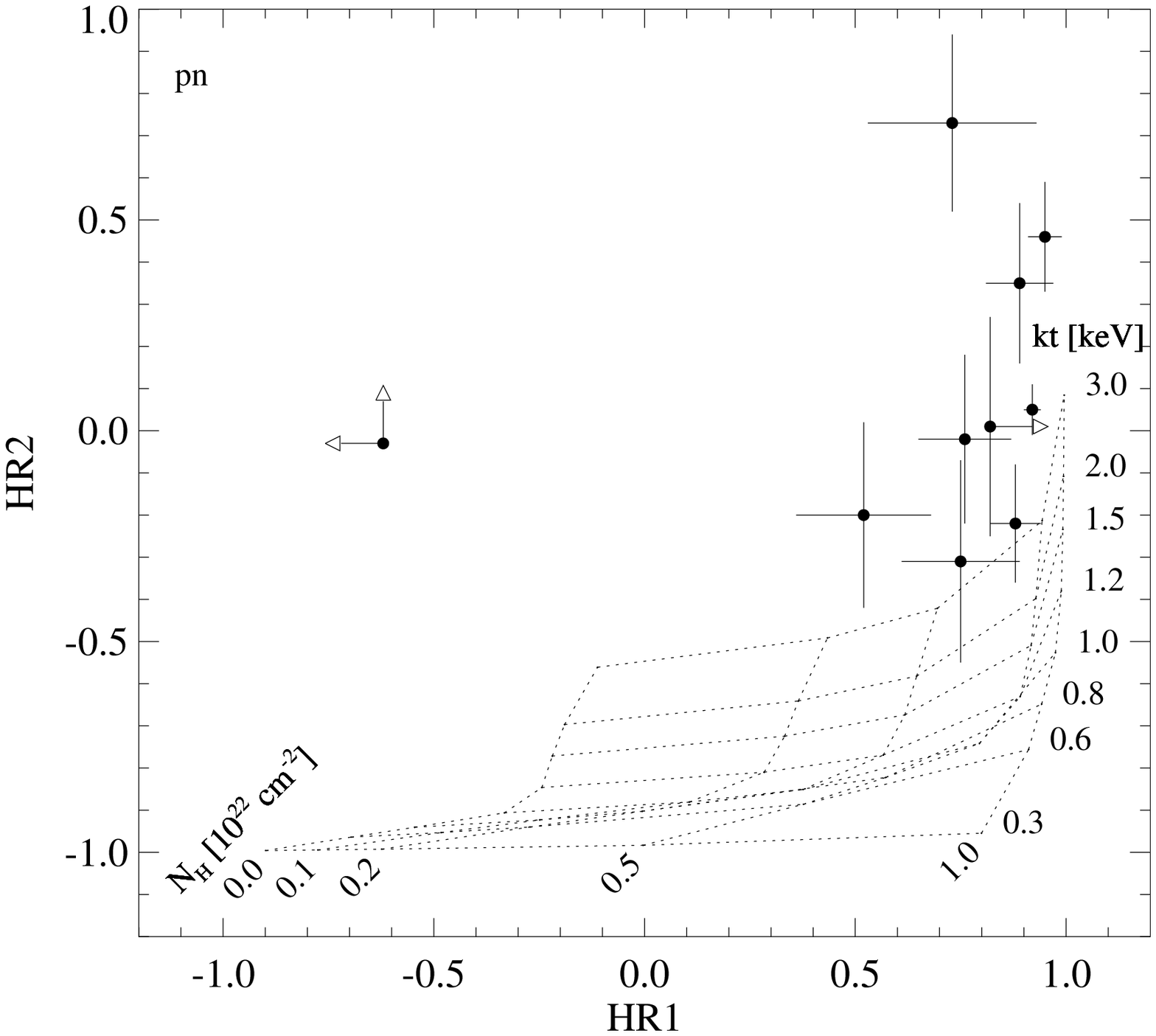}}
\resizebox{9cm}{!}{\includegraphics{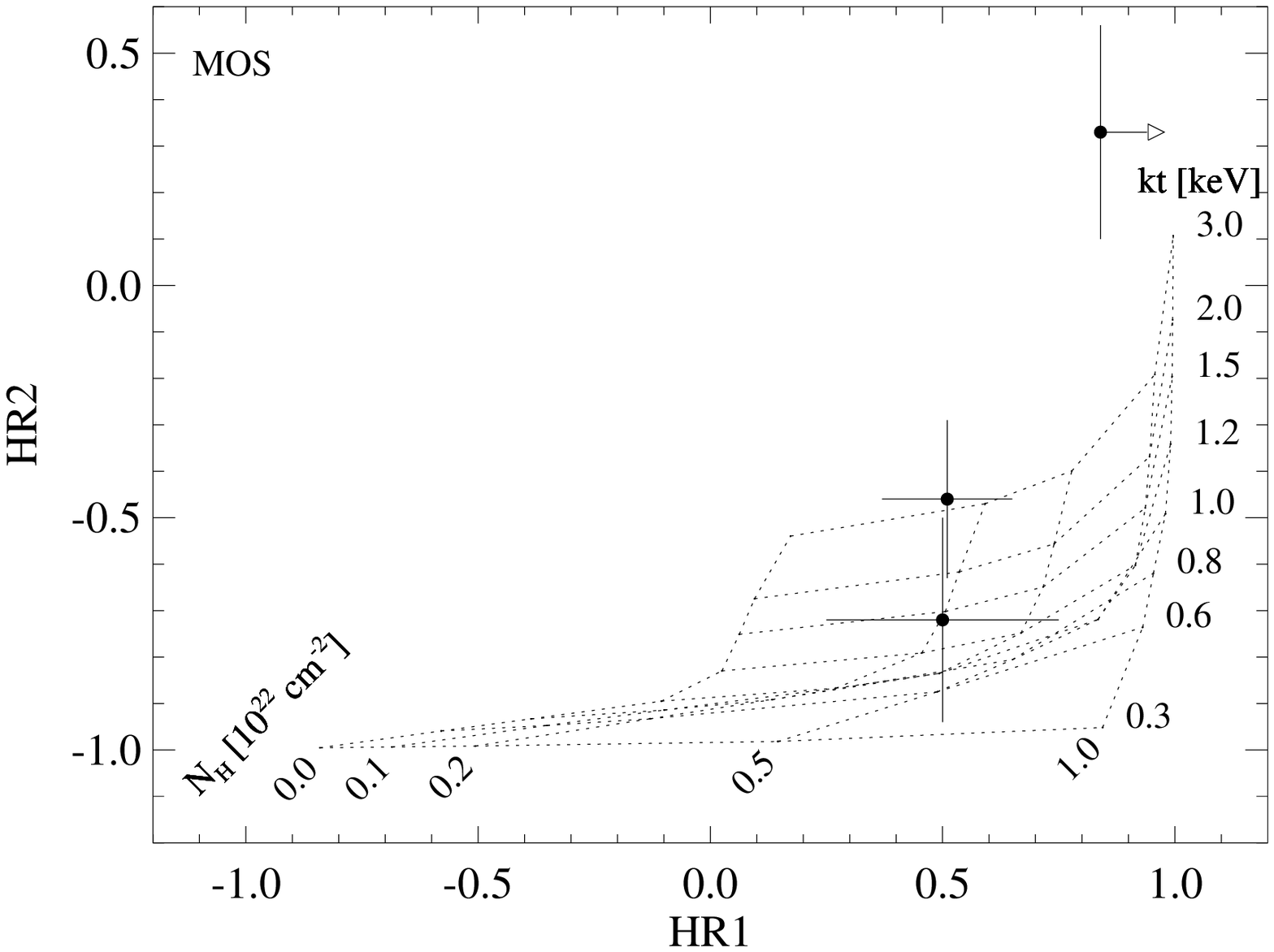}}
\caption{EPIC pn and MOS hardness ratios for unidentified X-ray sources overplotted on a grid of column density and temperature for a 1-T thermal model; see Fig.~\protect\ref{fig:hr1_hr2_iden} and text in Sect.~\protect\ref{subsect:kt} for more details. Most objects are harder than typical spectra for the bulk of Cha\,I stars, suggesting either very young and strongly absorbed cloud members or extragalactic origin (see text in Sect.~\ref{subsect:uiden}).}
\label{fig:hr_uniden}
\end{center}
\end{figure}

For the remaining unknown X-ray objects we examined the hardness ratios (Fig.~\ref{fig:hr_uniden}). 
Most of the objects cluster in a region in the upper right of the area 
spanned by the thermal model. The fact that they are harder than typical coronal emitters 
(compare Fig.~\ref{fig:hr1_hr2_iden}) 
may indicate that these are extragalactic sources. The typical flux level of the unidentified
objects is $\sim 10^{-14}\,{\rm erg/s/cm^2}$. According to the $\log{N} - \log{S}$
distribution presented by \citey{Hasinger01.1} a total of $\sim 20$ extragalactic sources is
expected at this flux-level within the {\em XMM-Newton} field, in rough agreement with the
number of unidentified X-ray sources. 

On the other hand two among the most highly extincted Cha\,I stars (HM\,16 and CCE\,98-32)
showed similar hardness ratios, and an X-ray spectrum with a very pronounced high-energy tail
(see Fig.~\ref{fig:hr1_hr2_iden} and Sect.~\ref{subsect:hd97048_group}).
Therefore we caution that several cloud members may still be hidden behind substantial 
extinction.  
\citey{Gomez03.1} argued on basis of the apparent fall-off of the mass function in 
Cha\,I that up to $\sim 100$ low-mass stars and brown dwarfs remain to be identified
in this cloud. 
Some of these stars may be among the IR candidate members presented in the literature, 
and among the unidentified {\em XMM-Newton} sources. 

\section{Comparison to {\em ROSAT}}\label{sect:rosat}

In search for long-term variability we compare the {\em XMM-Newton} observation of 
April 2002 to earlier X-ray observations of the same stars. 
The full EPIC FOV was covered 11\,yrs earlier in two pointings  
with the {\em ROSAT} PSPC. The results of the {\em ROSAT} observations were discussed by
F93, \citey{Neuhaeuser99.1}, and CNK00. 
In lack of statistics that would allow for a spectral analysis F93
derived the X-ray luminosities assuming a 1\,keV plasma seen through an absorbing 
column of $0.2 \times 10^{22}\,{\rm cm^{-2}}$ for {\em all} X-ray sources. 
LFH96 presented new optical photometry and spectroscopy, 
updated the list of Cha\,I members, and compiled $L_{\rm x}$ for this new list  
following the assumptions on the spectral shape by F93. 

\begin{figure}
\begin{center}
\resizebox{9cm}{!}{\includegraphics{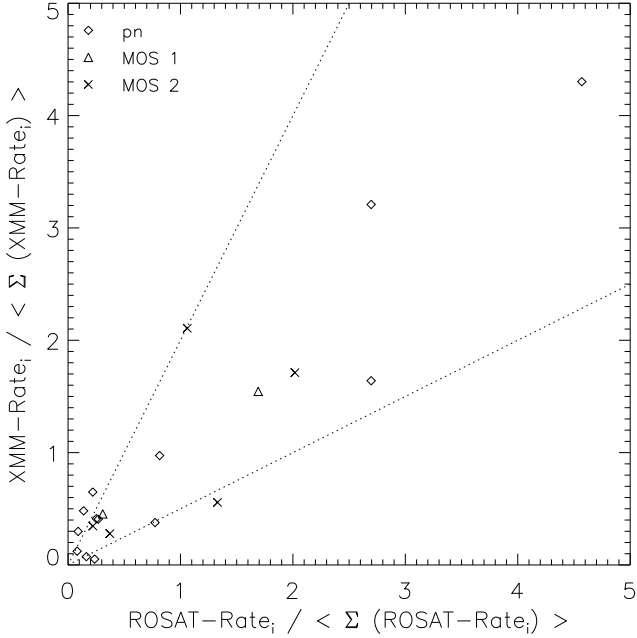}}
\caption{Relative X-ray brightness of stars in Cha\,I during the {\em XMM-Newton} and the {\em ROSAT} 
observation. Count rates refer to the $0.4-2.5$\,keV band. Only stars detected by both satellites are shown.
The dotted lines embrace variations by a factor two and less.} 
\label{fig:rate_rosat_xmm}
\end{center}
\end{figure}

The luminosities measured by F93 and LFH96 with {\em ROSAT} 
are systematically lower than those we derive from the {\em XMM-Newton}
spectra, in some cases by more than one order of magnitude. 
This is probably the effect of a 
serious underestimation of the absorption in the low-sensitivity {\em ROSAT} observations. 
In view of the difficulties in establishing luminosities based on simplified models
we prefer to compare the data on the count rate level.
We normalized the count rate of each X-ray source by the mean count rate of the total sample,
separately for {\em XMM-Newton} and {\em ROSAT}. This quantity characterizes the brightness
of the source with respect to the whole sample, and should remain approximately constant over
time. In Fig.~\ref{fig:rate_rosat_xmm} we confront the values derived from the two satellites 
with each other. The data scatters around the $1:1$ relation with little indications for
variations larger than a factor of two, signaling that there is no significant long-term 
variability. 
%
%

\section{Correlation with Stellar Parameters}\label{sect:corr}

The ratio of X-ray to bolometric luminosity is one of the most widely used activity 
indicators. Previous studies have repeatedly found 
that $L_{\rm x}/L_{\rm bol} \sim 10^{-3}$ for the most active stars, with a scatter
of $\sim 2$\,dex down to $L_{\rm x}/L_{\rm bol} \sim 10^{-5}$ for less active stars. 
This `saturation' is well-established on basis of {\em ROSAT} observations in various
star forming regions. In Cha\,I a linear regression to 
$\log{L_{\rm x}}$ versus $\log{L_{\rm bol}}$ provided a slope of $1.4 \times 10^{-4}$ 
(see F93). However, as a consequence of the 
underestimation of the X-ray luminosities of Cha\,I stars in {\em ROSAT} measurements, 
this mean $L_{\rm x}/L_{\rm bol}$ value as well as the saturation level must now 
be questioned. 

\begin{figure}
\begin{center}
\resizebox{9cm}{!}{\includegraphics{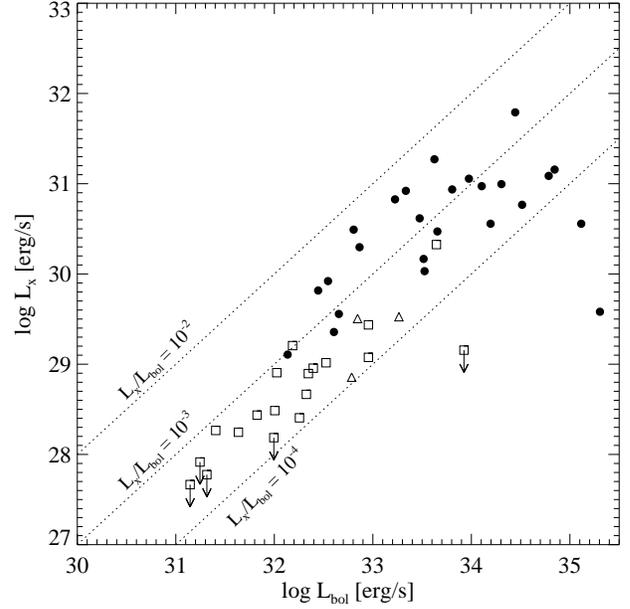}}
\caption{X-ray versus bolometric luminosity. 
Different plotting symbols denote different methods to derive $L_{\rm x}$: 
{\em filled circles} - Spectral fits with 2-T models and column density derived from $A_{\rm J}$, 
{\em open squares} - Assumed 1-T model with $\log{T} = 6.95$ and $A_{\rm J}$ from the literature, 
{\em open triangles} - Assumed 1-T model with $\log{T} = 6.95$ and $A_{\rm J}$ derived from the $JHK$ 
color-color diagram (Fig.~\ref{fig:jh_hk}).}
\label{fig:lx_lbol}
\end{center}
\end{figure}

We examine the $L_{\rm x}/L_{\rm bol}$ relation combining the {\em XMM-Newton}
data with the $L_{\rm bol}$ values derived in Sect.~\ref{subsect:lbol_age}.
The result is shown in Fig.~\ref{fig:lx_lbol}. The well-known correlation between 
$L_{\rm x}$ and $L_{\rm bol}$ is evident, with the common scatter of $\sim 2$\,dex. 
The net effect of our more realistic X-ray luminosities is shifting the correlation
by $\sim 1$\,dex upwards, such that nearly all stars end up with 
$L_{\rm x}/L_{\rm bol} > 10^{-4}$. 
Stars for which $L_{\rm x}$ has been determined from the spectrum (symbolized by filled
circles in Fig.~\ref{fig:lx_lbol}) tend to have high 
$L_{\rm x}/L_{\rm bol}$, with saturation near $10^{-2.5}$. 
For stars with too little signal for an analysis of the X-ray spectrum 
(open plotting symbols in Fig.~\ref{fig:lx_lbol}) $L_{\rm x}$ was derived {\it assuming} 
a spectral model, and the saturation level is found to be $\sim 10^{-3}$. 
The most obvious explanation is a bias introduced by the simplified model assumptions:  
In particular the assumption of an iso-thermal plasma and uniform extinction lead to an 
underestimate of the emission measure. As a consequence  
the true saturation level of the faint stars in Cha\,I could be similar to that of 
the brighter Cha\,I members. 
However, a physical effect, i.e. changing saturation level with stellar mass,
can not be excluded from the data. Indeed, this latter interpretation would be in accordance with 
a {\em Chandra} study of the Orion Nebula Cluster which suggested 
that $L_{\rm x}/L_{\rm bol}$ declines at the low-mass end of the MS (\cite{Flaccomio03.2}).

The two outliers in Fig.~\ref{fig:lx_lbol} with the highest bolometric luminosity 
are the intermediate mass stars HD\,97048 and KG\,2001-78, which may be subject to different
X-ray emission mechanisms. A similar trend was seen in a {\em Chandra} observation 
of the young NGC\,1333 cluster. \citey{Preibisch02.1} argued that a bifurcation seen 
for stars with high bolometric luminosity in the $L_{\rm x} - L_{\rm bol}$ diagram 
is produced by A-type stars with late-type companions. 
In our case the companion hypothesis is improbable: HD\,97048 is a well-studied single
star, and KG\,2001-78 shows an X-ray spectrum very distinct from that of a TTS. 
A large spread of $\log{(L_{\rm x}/L_{\rm bol})}$ was also reported for intermediate 
mass stars in Orion (\cite{Feigelson03.1}, \cite{Flaccomio03.2}), giving rise to the speculation
that, depending sensitively on the details of the internal structure, 
various emission mechanisms may be at work in this critical mass range. This view 
seems to be supported by our observation of just two, but apparently 
very different, A-type X-ray emitters in Cha\,I. 
%
%

We also examined the relation between X-ray luminosity and stellar age. 
Most of the stars are found in a small age range between $\sim 2-6$\,Myr. The near-to
coevality of the Cha\,I population and its implication on the star formation rate 
has been discussed by LFH96. The need for a strong increase of the star formation rate 
in the past can be avoided if the older population is incomplete, 
e.g. because the stars have dispersed and thus escaped recognition as a cloud member.  
Interestingly, three of the four undetected VLM ChaH$\alpha$ objects have ages $> 10$\,Myr. 
From the presently derived upper limits 
we can not say whether the dynamo of the lowest mass stars and brown dwarfs 
has completely shut off at this age,
or whether these objects just follow a trend of declining X-ray luminosity with age. 

It has been argued that the decrease of $L_{\rm x}$ with age may be an effect of the
loss of coupling between the increasingly neutral atmosphere and the magnetic
field (\cite{Mohanty02.1}). 
We find no evidence for a temperature related dynamo shut-off for the lowest masses: 
The undetected VLM objects have effective temperatures comparable to the detected VLM objects. 
For the whole sample a strong correlation between $L_{\rm x}$ and $T_{\rm eff}$ is evident.
However, this correlation disappears if $L_{\rm x}$ is replaced by 
$L_{\rm x}/L_{\rm bol}$. Therefore, the $L_{\rm x}-T_{\rm eff}$-correlation can be explained 
by the dependence of 
$L_{\rm bol}$ on $T_{\rm eff}$ together with the relatively small scatter of the stars in 
terms of $L_{\rm x}/L_{\rm bol}$. 

\begin{figure}
\begin{center}
\resizebox{9cm}{!}{\includegraphics{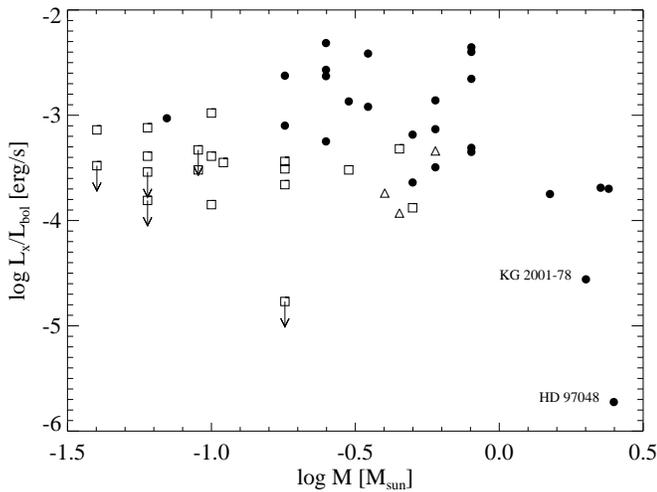}}
\caption{Ratio of X-ray to bolometric luminosity versus mass derived from the PMS models by 
\protect\citey{Baraffe98.1} and \protect\citey{Palla99.1}. 
The meaning of the plotting symbols is the same as in Fig.~\ref{fig:lx_lbol}.
Note the large spread for intermediate-mass which likely reflects directly the changes in the
interior structure.} 
\label{fig:lx_mass}
\end{center}
\end{figure}

Finally, the relation of X-ray emission with stellar mass is shown in Fig.~\ref{fig:lx_mass}.
This diagram shows that the spread in $L_{\rm x}/L_{\rm bol}$ increases dramatically for
intermediate-mass stars: The less evolved ones still possess substantial convection zones,
and display X-ray emission levels similar to lower-mass stars. KG\,2001-78 and HD\,97048 are
in their final approach to the MS, fully radiative, and consequently less active. 
A similar trend was found in the ONC by \citey{Flaccomio03.2} and \citey{Feigelson02.1}. 

\section{Summary}\label{sect:summary}

In a 30\,ksec {\em XMM-Newton} observation of the Cha\,I South cloud
we detected 
the intermediate-mass HAeBe star HD\,97048, 
all known TTS members except Sz\,23 (confused with VW\,Cha) and the IR\,Nebula, 
most of the VLM H$\alpha$ objects near or below the substellar limit, 
and three photometric Cha\,I candidates two of which are probably late-type stars
and one is likely to be an A-type star. 
A possibly sub-Jupiter mass object recently proposed as a Cha\,I candidate by
\citey{Comeron04.2} is in the central part of the {\em XMM-Newton} field,
but not detected. 

Our X-ray detection of candidate young stars identified by means of 
NIR photometry is an important indication for them being true members of the star 
forming region. We argue on basis of their X-ray properties that some additional new 
cloud members may be among the unidentified X-ray sources. 
The nature of these objects as well as that of the detected NIR candidate 
members will be revealed in optical/IR follow-up studies. 

We performed a detailed spectral analysis of the brighter half of the Cha\,I members
in the {\em XMM-Newton} field. The spectra are described by thermal emission from a hot,
optically thin plasma, and a photo-absorption term taking account of interstellar and/or
circumstellar extinction. Absorbing column densities derived from {\em Chandra} and 
{\em XMM-Newton} spectra were used by 
\citey{Vuong03.1} to constrain the gas-to-dust extinction relation. Their analysis of X-ray data
from various star forming regions showed that in $\rho$\,Oph the measured $N_{\rm H,X}$ 
are systematically lower than expected from $A_{\rm J}$ assuming the standard value of 
$N_{\rm H}/A_{\rm V} = (1.8-2.2)\,10^{21}\,{\rm cm^{-2}}$ per magnitude; see references in 
\citey{Vuong03.1}. 
Their sample in Cha\,I comprised only 4 reasonably X-ray bright stars in the northern
cloud observed with {\em Chandra}, that span a small range in optical extinction.  
We point out that low absorption is not a general characteristic of the Cha\,I cloud. 
In fact, the {\em XMM-Newton} FOV contains some objects with $A_{\rm V} > 10$\,mag. But the
statistics in their X-ray spectra are too small to derive a well-constrained $N_{\rm H}$. 
We consider the spectral fits with column density fixed on the value expected from the
canonical $N_{\rm H} - A_{\rm V}$ relation more reliable, and 
feel that the interpretation of the $N_{\rm H} - A_{\rm V}$ relation based on present-day X-ray 
observations demands caution. 

The X-ray temperatures of the coronal sources in Cha\,I are similar to those found by
\citey{Favata03.1} in the Taurus cloud L\,1551, but lower than those found in recent
X-ray observations of IC\,348 (\cite{Preibisch02.1}), NGC\,1333 (\cite{Getman02.1}),
and the Orion Nebula Cloud (\cite{Feigelson02.1}). 
This can probably be attributed to the use of different
instruments and model assumptions. The latter three studies are based on {\em Chandra} 
and adopted a 1-T approach.
It is a well-known fact that 1-T models do not represent a valid description of the
temperature structure in stellar coronae. Nevertheless, they are often used to describe
low-resolution X-ray spectra with low numbers of counts. 
We find that 1-T models have a tendency to underestimate $L_{\rm x}$ because they
underestimate absorption. 
For this reason we chose to describe all Cha\,I members which are bright enough
for spectral analysis by a 2-T model. 
Not surprisingly then our spectral analysis brought forth systematically higher X-ray 
luminosities as compared to the previous estimates derived from {\em ROSAT} observations. 
The X-ray luminosities of the low-mass stars in Cha\,I have now been re-adjusted,
and reveal a saturation level near $10^{-2.5...-3}$ (versus $\sim 10^{-4}$ suggested
by {\em ROSAT}). 
On basis of this conclusion is it impractical to engage in a study of long-term
variability based on X-ray luminosities. We have, however, 
no signs for variability exceeding a factor of $\sim 2$ within the last $11$\,yrs. 

The confirmation of X-ray emission from all but one of the {\em ROSAT} 
detected VLM ChaH$\alpha$ objects provides important support for the reliability
of the {\em ROSAT} source detection process. 
ChaH$\alpha$\,7 is the first M8 brown dwarf in Cha\,I detected in X-rays. 
In addition, the higher sensitivity and continuous data stream of {\em XMM-Newton} 
has allowed for the first time to examine the spectral characteristics and the time 
variability of the latest type stars and brown dwarfs in Cha\,I. 
We find no evidence for a dramatic change in the X-ray
properties (such as X-ray temperature and X-ray luminosity) at the substellar limit. 
The oldest VLM objects are undetected presumably due to the decline of X-ray luminosity
with age, and not to an effect of the atmospheric temperature. 
In terms of variability the lowest mass Cha\,I members behave similar 
to higher-mass TTS in the cloud. In particular they are shown to undergo flares.

\begin{acknowledgements}

BS acknowledges financial support from the European Union by the Marie
Curie Fellowship Contract No. HPMD-CT-2000-00013. BS wants to thank F. Comer\'on
and J. Alves for discussions on the IR properties of young stars, 
and K. Briggs and I. Pillitteri for discussions on the data analysis. 
We thank the referee, F. Comer\'on, for very careful reading and useful
comments on essential points. 
{\em XMM-Newton} is an ESA science mission with instruments and contributions directly
funded by ESA Member States and the USA (NASA).

\end{acknowledgements}


\begin{thebibliography}{}

\bibitem[\protect\astroncite{Anders \& Grevesse}{1989}]{Anders89.1}
Anders E. \& Grevesse N., 1989, Geochimica et Cosmochimica Acta 53, 197

\bibitem[\protect\astroncite{Bailey}{1998}]{Bailey98.1}
Bailey J., 1998, MNRAS 301, 161

\bibitem[\protect\astroncite{Baldi et~al.}{2002}]{Baldi02.1}
Baldi A., Molendi S., Comastri A., et al., 2002, ApJ 564, 190

\bibitem[\protect\astroncite{Baraffe et~al.}{1998}]{Baraffe98.1}
Baraffe I., Chabrier G., Allard F. \& Hauschildt P. H., 1998, 
A\&A 337, 403

\bibitem[\protect\astroncite{Baud et~al.}{1984}]{Baud84.1}
Baud B., Beintema D. A., Wesselius P. R., et al., 1984, ApJ 278, L53 

\bibitem[\protect\astroncite{Bergh\"ofer et~al.}{1996}]{Berghoefer96.1}
Bergh\"ofer T. W., Schmitt J. H. M. M. \& Cassinelli J. P.,
1996, A\&A 118, 481

\bibitem[\protect\astroncite{Bessell \& Brett}{1988}]{Bessell88.1}
Bessell M. S. \& Brett J. M., 1988, PASP 100, 1134

\bibitem[\protect\astroncite{Brandeker et~al.}{2001}]{Brandeker01.1}
Brandeker A., Liseau R. Artymowicz P. \& Jayawardhana R., 2001, ApJ 561, L199

\bibitem[\protect\astroncite{Cambr\'esy et~al.}{1998}]{Cambresy98.1}
Cambr\'esy L., Copet E., Epchtein N., et al., 1998, A\&A 338, 977

\bibitem[\protect\astroncite{Carpenter et~al.}{2002}]{Carpenter02.1}
Carpenter J. M., Hillenbrand L. A., Skrutskie M. F. \& Meyer M. R., 2002, AJ 124, 1001

\bibitem[\protect\astroncite{Chelli et~al.}{1988}]{Chelli88.1}
Chelli A., Zinnecker H., Cruz-Gonzalez I., Carrasco L. \& Perrier C.,
1988, A\&A 207, 46

\bibitem[\protect\astroncite{Cohen \& Schwartz}{1984}]{Cohen84.1}
Cohen M. \& Schwartz R. D., 1984, AJ 89, 277

\bibitem[\protect\astroncite{Comer\'on \& Claes}{2004}]{Comeron04.2}
Comer\'on F. \& Claes P. 2004, ApJ 602, 298

\bibitem[\protect\astroncite{Comer\'on et~al.}{2004}]{Comeron04.1}
Comer\'on F., Reipurth B., Henry A. \& Fern\'andez M., 2004, A\&A in press 

\bibitem[\protect\astroncite{Comer\'on et~al.}{2003}]{Comeron03.1}
Comer\'on F., Fern\'andez M., Baraffe I., Neuh\"auser R. \& Kaas A. A., 2003, A\&A 406, 1001

\bibitem[\protect\astroncite{Comer\'on et~al.}{1999}]{Comeron99.1}
Comer\'on F., Rieke G. H. \& Neuh\"auser R., 1999, A\&A 343, 477

\bibitem[\protect\astroncite{Comer\'on et~al.}{2000}]{Comeron00.1}
Comer\'on F., Neuh\"auser R. \& Kaas A. A., 2000, A\&A 359, 269 
(CNK00)

\bibitem[\protect\astroncite{Corporon \& Lagrange}{1999}]{Corporon99.1}
Corporon P. \& Lagrange A.-M., 1999, A\&AS 136, 429

\bibitem[\protect\astroncite{D'Antona \& Mazzitelli}{1997}]{DAntona97.1}
D'Antona F. \& Mazzitelli I., 1997, MemSAI 68, 807

\bibitem[\protect\astroncite{den Herder et~al.}{2001}]{denHerder01.1}
den Herder, J. W., Brinkman, A. C., Kahn, S. M., et al. 2001, A\&A, 365, L7  

\bibitem[\protect\astroncite{Favata et~al.}{2003}]{Favata03.1}
Favata F., Giardino G., Micela G., Sciortino S. \& Damiani F., 2003, A\&A 403, 187

\bibitem[\protect\astroncite{Feigelson et~al.}{2003}]{Feigelson03.1}
Feigelson E. D., Gaffney J. A., Garmire G., et al., 2003, ApJ 584, 911

\bibitem[\protect\astroncite{Feigelson et~al.}{2002}]{Feigelson02.1}
Feigelson E. D., Broos P., Gaffney J. A., et al., 2002, ApJ 574, 258

\bibitem[\protect\astroncite{Feigelson et~al.}{1993}]{Feigelson93.1}
Feigelson E. D., Casanova S., Montmerle T. \& Guibert J., 1993, ApJ 416, 623
(F93)

\bibitem[\protect\astroncite{Feigelson \& Kriss}{1989}]{Feigelson89.1}
Feigelson E. D. \& Kriss G. A., 1989, ApJ 338, 262

\bibitem[\protect\astroncite{Flaccomio et~al.}{2003a}]{Flaccomio03.1}
Flaccomio E., Micela G. \& Sciortino S., 2003a, A\&A 397, 611

\bibitem[\protect\astroncite{Flaccomio et~al.}{2003b}]{Flaccomio03.2}
Flaccomio E., Damiani F., Micela G., et al., 2003b, ApJ 582, 398

\bibitem[\protect\astroncite{Gauvin \& Strom}{1992}]{Gauvin92.1}
Gauvin L. S. \& Strom K. M., 1992, ApJ 385, 217

\bibitem[\protect\astroncite{Gehrels}{1986}]{Gehrels86.1}
Gehrels N., 1986, ApJ 303, 336

\bibitem[\protect\astroncite{Getman et~al.}{2002}]{Getman02.1}
Getman K. V., Feigelson E. D., Townsley L., et al., 2002, ApJ 575, 354

\bibitem[\protect\astroncite{Ghez et~al.}{1997}]{Ghez97.1}
Ghez A. M., McCarthy D. W., Patience J. L. \& Beck T. L., 1997, ApJ 481, 378

\bibitem[\protect\astroncite{G\'omez \& Mardones}{2003}]{Gomez03.1}
G\'omez M. \& Mardones D., 2003, AJ 125, 2134

\bibitem[\protect\astroncite{G\'omez \& Kenyon}{2001}]{Gomez01.1}
G\'omez M. \& Kenyon S., 2001, AJ 121, 974

\bibitem[\protect\astroncite{Hamaguchi et~al.}{2000}]{Hamaguchi00.1}
Hamaguchi K., Terada H., Bamba A. \& Koyama K., 2000, ApJ 532, 1111

\bibitem[\protect\astroncite{Hartigan}{1993}]{Hartigan93.1}
Hartigan P., 1993, AJ 105, 1511

\bibitem[\protect\astroncite{Hartmann et~al.}{1997}]{Hartmann97.1}
Hartmann L., Cassen P. \& Kenyon S. J., 1997, ApJ 475, 770

\bibitem[\protect\astroncite{Hasinger et~al.}{2001}]{Hasinger01.1}
Hasinger G., Altieri B., Arnaud M., et al., 2001, A\&A 364, L45

\bibitem[\protect\astroncite{Henize \& Mendoza}{1973}]{Henize73.1}
Henize K. G. \& Mendoza E. E., 1973, ApJ 180, 180

\bibitem[\protect\astroncite{Huenemoerder et~al.}{1994}]{Huenemoerder94.1}
Huenemoerder D., Lawson W., A. \& Feigelson E. D., 1994, MNRAS 271, 967 

\bibitem[\protect\astroncite{Imanishi et~al.}{2001}]{Imanishi01.1}
Imanishi K., Tsujimoto M. \& Koyama K., 2001, ApJ 563, 361

\bibitem[\protect\astroncite{Jansen et~al.}{2001}]{Jansen01.1}
Jansen, F., Lumb, D., Altieri, B., et al. 2001, A\&A, 365, L1 

\bibitem[\protect\astroncite{Kenyon \& G\'omez}{2001}]{Kenyon01.1}
Kenyon S. \& G\'omez M., 2001, AJ 121, 2673

\bibitem[\protect\astroncite{Kenyon \& Hartmann}{1995}]{Kenyon95.1}
Kenyon S. \& Hartmann L., 1995, ApJS 101, 117

\bibitem[\protect\astroncite{Kraft et~al.}{1991}]{Kraft91.1}
Kraft R. P., Burrows D. N. \& Nousek J. A., 1991, ApJ 374, 344

\bibitem[\protect\astroncite{Lawson et~al.}{1996}]{Lawson96.1}
Lawson W. A., Feigelson E. D. \& Huenemoerder D. P., 1996, MNRAS 280, 1071
(LFH96)

\bibitem[\protect\astroncite{L\'opez~Mart\'\i ~et~al.}{2004}]{LopezMarti04.1}
L\'opez Mart\'\i~B., Eisl\"offel J., Scholz A. \& Mundt R., 2004, A\&A 416, 555

\bibitem[\protect\astroncite{Luhman}{1999}]{Luhman99.1}
Luhman K. L., 1999, ApJ 525, 466

\bibitem[\protect\astroncite{Luhman \& Rieke}{1998}]{Luhman98.1}
Luhman K. L. \& Rieke G. H., 1998, ApJ 497, 354

\bibitem[\protect\astroncite{Mart\'\i n \& Bouy}{2002}]{Martin02.1}
Mart\'\i n E. L. \& Bouy H., 2002, New Astron. 7, 595

\bibitem[\protect\astroncite{Mart\'\i n}{1998}]{Martin98.1}
Mart\'\i n E. L., 1998, AJ 115, 351

\bibitem[\protect\astroncite{Mewe et~al.}{1985}]{Mewe85.1}
Mewe R., Gronenschild E. H. B. M. \& van den Oord G. H. J., 1985, A\&AS 62, 197

\bibitem[\protect\astroncite{Mewe et~al.}{1995}]{Mewe95.1}
Mewe R., Kaastra J., S., Schrijver C. J., van den Oord G. H. J. \& Alkemade
F. J. M., 1995, A\&A 296, 477

\bibitem[\protect\astroncite{Mohanty et~al.}{2002}]{Mohanty02.1}
Mohanty S., Basri G., Shu F., Allard F. \& Chabrier G., 2002, ApJ 571, 469

\bibitem[\protect\astroncite{Morrison \& McCammon}{1983}]{Morrison83.1}
Morrison R. \& McCammon D., 1983, ApJ 270, 119

\bibitem[\protect\astroncite{Nakajima et~al.}{2003}]{Nakajima03.1}
Nakajima H., Imanishi K., Takagi S.-I., Koyama K. \& Tsujimoto M., 2002, 
PASJ 55, 635

\bibitem[\protect\astroncite{Neuh\"auser et~al.}{2003}]{Neuhaeuser03.1}
Neuh\"auser R., Guenther E. W. \& Brandner W., 2003, In: Brown Dwarfs, IAU Symp.~211,
Mart\'\i n E. L. (ed.), Astronomical Society of the Pacific, 309

\bibitem[\protect\astroncite{Neuh\"auser et~al.}{2002}]{Neuhaeuser02.1}
Neuh\"auser R., Brandner W., Alves J., Joergens V. \& Comer\'on F., 
2002, A\&A 384, 999

\bibitem[\protect\astroncite{Neuh\"auser et~al.}{1999}]{Neuhaeuser99.1}
Neuh\"auser R., Brice\~no C., Comer\'on F., et al., 1999, A\&A 343, 883

\bibitem[\protect\astroncite{Neuh\"auser \& Comer\'on}{1998}]{Neuhaeuser98.1}
Neuh\"auser R. \& Comer\'on F., 1998, Science 282, 83 

\bibitem[\protect\astroncite{Oasa et~al.}{1999}]{Oasa99.1}
Oasa Y., Tamura M., Sugitani K., 1999, ApJ 526, 336

\bibitem[\protect\astroncite{Palla \& Stahler}{1999}]{Palla99.1}
Palla F. \& Stahler S. W., 1999, ApJ 525, 772

\bibitem[\protect\astroncite{Paresce et~al.}{1984}]{Paresce84.1}
Paresce F., 1984, AJ 89, 1022

\bibitem[\protect\astroncite{Perlman et~al.}{1998}]{Perlman98.1}
Perlman E. S., Padovani P., Giommi P., et al., 1998, AJ 115, 1253

\bibitem[\protect\astroncite{Persi et~al.}{2000}]{Persi00.1}
Persi P., Marenzi A. R., Olofsson G., et al., 2000, A\&A 357, 219

\bibitem[\protect\astroncite{Preibisch}{2003}]{Preibisch03.1}
Preibisch Th., 2003, A\&A 401, 543

\bibitem[\protect\astroncite{Preibisch \& Zinnecker}{2002}]{Preibisch02.1}
Preibisch Th. \& Zinnecker H., 2002, AJ 123, 1613

\bibitem[\protect\astroncite{Preibisch}{1997}]{Preibisch97.1}
Preibisch Th., 1997, A\&A 320, 525

\bibitem[\protect\astroncite{Prusti et~al.}{1992}]{Prusti92.1}
Prusti T., Whittet D. C. B. \& Wesselius P. R., 1992, MNRAS 254, 361

\bibitem[\protect\astroncite{Raymond \& Smith}{1977}]{Raymond77.1}
Raymond J. C. \& Smith B. W., 1977, ApJS 35, 419

\bibitem[\protect\astroncite{Rieke \& Lebofsky}{1985}]{Rieke85.1}
Rieke G. H. \& Lebofsky M. J., 1985, ApJ 288, 618 (RL85)

\bibitem[\protect\astroncite{Rutledge et~al.}{2000}]{Rutledge00.1}
Rutledge R. E., Basri G., Mart\'\i n E. L. \& Bildsten L., 2000, 
ApJ 538, L141

\bibitem[\protect\astroncite{Schmidt-Kaler}{1982}]{SchmidtKaler82.1}
Schmidt-Kaler T., 1982, In: Landolt-B\"ornstein, Group VI, Vol.2, 
K.-H. Hellwege (ed.), Berlin, Springer, 454

\bibitem[\protect\astroncite{Schwartz}{1977}]{Schwartz77.1}
Schwartz R. D., 1977, ApJS 35, 161

\bibitem[\protect\astroncite{Skinner et~al.}{1997}]{Skinner97.1}
Skinner S. L., G\"udel M., Koyama K. \& Yamauchi S, 1997, ApJ 486, 886

\bibitem[\protect\astroncite{Skinner et~al.}{1993}]{Skinner93.1}
Skinner S. L., Brown A. \& R. T. Stewart, 1993, ApJS 87, 217

\bibitem[\protect\astroncite{Stelzer \& Hu\'elamo}{2000}]{Stelzer00.2}
Stelzer B. \& Hu\'elamo N., 2000, A\&A 363, 667

\bibitem[\protect\astroncite{Stelzer et~al.}{2000}]{Stelzer00.1}
Stelzer B., Neuh\"auser R. \& Hambaryan V., 2000, A\&A 356, 949

\bibitem[\protect\astroncite{Str\"uder et~al.}{2001}]{Strueder01.1}
Str\"uder L., Briel U. G., Dennerl K., et al., 2001, A\&A 365, L18  

\bibitem[\protect\astroncite{Takami et~al.}{2003}]{Takami03.1}
Takami M., Bailey J. \& Chrysostomou A., 2003, A\&A 397, 675

\bibitem[\protect\astroncite{Tozzi et~al.}{2001}]{Tozzi01.1}
Tozzi P., Rosati P., Nonino M., et al., 2001, ApJ 562, 42

\bibitem[\protect\astroncite{Tsuboi et~al.}{1998}]{Tsuboi98.1}
Tsuboi Y., Koyama K., Murakami H., et al., 1998, ApJ 503, 894 

\bibitem[\protect\astroncite{Turner et~al.}{2001}]{Turner01.1}
Turner, M. J. L., Abbey, A., Arnaud, M., et al. 2001, A\&A, 365, L27  

\bibitem[\protect\astroncite{van den Ancker et~al.}{1998}]{vandenAncker98.1}
van den Ancker M. E., de Winter D. \& Tjin A Djie H. R. E., 1998, A\&A 330, 145

\bibitem[\protect\astroncite{Vuong et~al.}{2003}]{Vuong03.1}
Vuong M. H., Montmerle T., Grosso N., et al., 2003, A\&A 408, 581

\bibitem[\protect\astroncite{Walter}{1992}]{Walter92.1}
Walter F. M., 1992, AJ 104, 758

\bibitem[\protect\astroncite{Whittet et~al.}{1997}]{Whittet97.1}
Whittet D. C. B., Prusti T., Franco G. A. P., et al., 1997, A\&A 327, 1194

\bibitem[\protect\astroncite{Wichmann et~al.}{1998}]{Wichmann98.1}
Wichmann R., Bastian U., Krautter J., Jankovics I. \& Rucinski S. M., 1998, MNRAS 301, L39

\bibitem[\protect\astroncite{Wichmann et~al.}{1997}]{Wichmann97.1}
Wichmann R., Krautter J., Covino E., et al., 1997, A\&A 320, 185

\bibitem[\protect\astroncite{Zapatero et~al.}{1997}]{Zapatero97.1}
Zapatero Osorio M. R., Mart\'\i n E. L. \& Rebolo R., 1997, A\&A 323, 105

\bibitem[\protect\astroncite{Zinnecker \& Preibisch}{1994}]{Zinnecker94.1}
Zinnecker H. \& Preibisch T., 1994, A\&A 292, 152

\end{thebibliography}
\end{document}